\documentclass[12pt, a4paper]{article}
\usepackage{graphicx} 
\usepackage[a4paper, margin=20mm]{geometry}
\usepackage{amsmath}
\usepackage{authblk}
\usepackage{textcomp, gensymb}
\usepackage{float}
\usepackage{xcolor}

\usepackage[style=chem-acs]{biblatex}
\usepackage{setspace}
\newcommand{\um}{{\textmu}m}

\title{Multifunctional Imaging with an Inverse-Designed Nonlocal Metasurface}
\author[1, *]{Lincoln Clark}
\author[2]{Yang Xu}
\author[3]{Cat-Uyen Phan}
\author[4]{David J. Byrne}
\author[5, 6]{Shikun Ma}
\author[1]{Lukas Wesemann}
\author[5]{Elizabeth Hinde}
\author[3]{Kylie L. Gorringe}
\author[1]{Ann Roberts}
\date{}
\affil[1]{ARC Centre of Excellence for Transformative Meta-Optical Systems, School of Physics, University of Melbourne, Victoria 3010, Australia}
\affil[2]{ARC Centre of Excellence for Transformative Meta-optical Systems, Department of Electrical and Electronic Engineering, University of Melbourne, Victoria 3010, Australia}
\affil[3]{The Sir Peter MacCallum Dept of Oncology, University of Melbourne, Victoria 3010, Australia}
\affil[4]{Department of Pathology, Peter MacCallum Cancer Centre, Melbourne, Victoria 3000, Australia}
\affil[5]{School of Physics, The University of Melbourne, Victoria 3010, Australia}
\affil[6]{Department of Biochemistry and Pharmacology, The University of Melbourne, Victoria 3010, Australia}
\affil[*]{Corresponding author: clarklj@student.unimelb.edu.au}

\begin{document}

\maketitle


\begin{refsection}

\section{Abstract}

Nonlocal metasurfaces enable all-optical processing of spatial information in optical fields. Here, we demonstrate a topology-optimised metasurface that switches between phase-contrast and brightfield imaging modalities via polarisation control, eliminating the need to reposition optical components or use computational techniques to image transparent samples. Specifically, for one polarisation state, an asymmetric transfer function about normal incidence performs a first order derivative on the phase, producing pseudo-3D phase-contrast images of transparent biological samples while the orthogonal state returns the result of the identity operator. This work extends inverse-design methods to reconfigurable phase-contrast microscopy and quantitative analogue optical computation with applications in biological imaging, medical diagnostics, and materials characterisation.

\subsection*{Keywords}

Topology Optimisation, Metasurfaces, Image Processing, Optical Computation, Inverse Design

\section{Introduction}

Metasurfaces are an emerging technology capable of performing real-time image processing and computational operations in an ultra-compact form factor \cite{ZangenehNejad2020}. The enhancement and extraction of features in an image is a key computational step in augmented reality, object recognition and machine vision. Performing these operations digitally leads to power consumption, transmission bandwidth, and latency costs, which can be reduced by performing these operations prior to detection directly with light, where computation occurs inherently at light-speed. Furthermore, analogue computational operations with light can access additional information in the optical field such as phase and polarisation, which conventional image detection systems typically discard. Phase in particular, is important in metrology, wavefront sensing, and imaging of unstained biological samples. Non or weakly absorbing objects, such as cells or tissue samples exhibit poor contrast in bright field microscopy. Refractive index and thickness variations in the specimens, however, modulate the phase of the transmitted optical field. By being able to image the phase, we can collect information about the morphology and other properties of these transparent samples. As a consequence, methods such as Zernike phase contrast \cite{Zernike1942, Zernike1942_2} or differential interference contrast (DIC) microscopy \cite{lang_1982} were developed to turn this phase variation into a visible intensity modulation, but generally at the expense of system size, complexity and cost. In such systems, switching to an alternative imaging mode requires additional optical components to be introduced into or removed from the system. Computational techniques can also be used to image and recover the phase of a sample, such as using the transport of intensity equation \cite{Streibl1984, Zuo2020} however this requires capturing several images, requiring either mechanical movement or complex optical systems and subsequent image registration for single shot acquisition. Other methods such as digital holographic microscopy \cite{Monemhaghdoust2014} and ptychography \cite{Wang2025Ptch} require complex optical setups and/or intensive computation. There remains an ongoing need to develop ultra-compact real-time imaging systems that can perform multiple image processing modalities directly with light. These all optical image processing systems permit direct computation of phase information, without the need to utilise additional optical components or moving parts in a microscope to change the imaging modality. 

Spatial filtering is a well-known technique for optically processing images, where filtering is performed in momentum space. Using the Fourier transform property of a lens, a spatial filter can be implemented by introducing a mask with a spatially varying complex transmission function into the Fourier plane \cite{Goodman1996}, in what is known as a $4f$-filtering setup. This approach, however, requires extra optical components and macroscopic propagation distances. To significantly reduce the axial extent of the system by many orders of magnitude, we can use an object plane approach \cite{Case1979}. Since the momentum of a plane wave is related to the direction it travels, by designing a filter with a transmission that depends on the plane wave's angle of incidence, we can perform spatial filtering directly in the object or image plane with no extra optical components or propagation distances \cite{wesemann2021meta}. Several of these ``non-local" devices that perform edge detection in the object plane have been previously demonstrated using thin film stacks \cite{Zhu2017, Wesemann2019}, gratings \cite{Cordaro2019}  and metasurfaces \cite{Komar2021}. This object plane spatial filtering approach has also been applied to phase imaging using resonant waveguide gratings \cite{Wesemann2021NEC, Ji2022}, and arrangements of plasmonic nanobars \cite{Davis2019, Wesemann2022asym, haiwei_phase}.

These works describe metasurfaces that perform only a single imaging operation. To simplify optical systems and reduce the need to move optical components in and out of the system to change the image processing operation, we require metasurfaces that can perform multiple functionalities. Currently, there are several promising mechanisms to encode this flexibility into a metasurface \cite{Nemati2018, Abdelraouf2022} to enable reconfigurable behaviour post-fabrication. The refractive index of materials of which the structure is composed can be modified through a change in phase of a constituent material \cite{Earl2016, Oguntoye2023, Liu2020, Murali2024}, using liquid crystals \cite{Izdebskaya2023} or chemical actuation \cite{Karst2021, Karst2022, Wang2023}. The geometry of the metasurface can be modified though mechanical stretching or micro-electromechanical systems \cite{Zhao2024}. Finally, a static metasurface design can be used and the type of illumination changed, such as wavelength \cite{Wang2016}, polarisation \cite{Arbabi2015, Xiong2023}, angle of incidence \cite{Kamali2017} and spatial mode \cite{Fang2019}. Although there are several promising mechanisms to design multifunctional metasurfaces, the application to image processing is an emerging area. There have been several demonstrations of nonlocal metasurfaces that can switch from brightfield to edge-detection modalities using various tuning mechanisms. For example, using flexible substrates in combination with mechanical stretching \cite{Zhang2021stretch}, water sensitive hydrogels \cite{Dai2023gel} and phase change materials, VO$_2$ \cite{Cotrufo2024vo2} and Sb$_2$Se$_3$ \cite{Yang2025sb2se3}. In the mid-infrared, free-form topology optimised metasurfaces have demonstrated polarisation and wavelength switchable image processing \cite{Pearson2025}. Although there have been several works focusing on switching from a brightfield to an edge detection mode, the application to phase imaging in particular has been limited.

Metasurface designs for multifunctional phase imaging have been hindered by lack of compact designs and mechanical movements, lossy materials or off-normal operation. Switchable brightfield, differential and quantitative phase imaging was demonstrated in a $4f$ system with a metasurface filter placed in the Fourier plane \cite{Yang2026dif}. Further to the constraints described above, this approach achieved tuning by mechanically shifting the metasurface position. Kwon et al. demonstrated that using two cascaded metasurfaces and a polarisation sensitive camera, quantitative phase gradients were able to be extracted from phase objects \cite{Kwon2019qpgm} enabling imaging of transparent samples, but the use of dual metasurfaces complicates the fabrication process. Meanwhile, a non-local metasurface using wavelength and polarisation multiplexing consisting of silver bars on a high refractive index layer was able to achieve quantitative phase imaging in a compact form factor \cite{haiwei_phase}. By changing the wavelength, the metasurface can switch from performing differential phase contrast along the $x$ to $y$ axis. The use of plasmonic resonances leads to loss and affects the imaging contrast. As for low-loss dielectric metasurfaces, a metasurface consisting of rectangular meta-atoms demonstrated polarisation switchable brightfield/phase-imaging \cite{Sulejman2025}. The rectangular meta-atoms creates a polarisation dependence in the excitation of Mie-resonances, leading to switchable behaviour. This metasurface, however, was required to be tilted with respect to the optical axis which limits integration into ultra-compact imaging systems. 

Here we use topology optimisation to design a dielectric metasurface that can perform switchable phase contrast/brightfield imaging at normal incidence. The topology optimisation algorithm explores a very large design space to iteratively determine the distribution of material in the metasurface to achieve a desired optical response, with minimal constraints on the geometries produced. Rather than optimising the parameters of some candidate geometry, the radius of a dielectric cylinder, for example, we optimise the spatial distribution of permittivity in the metasurface. Topology optimisation has been used previously to design ultra-high efficiency metasurfaces for a variety of applications, for example high efficiency plasmonic nanotweezers \cite{Nelson2023TOtweezers}, enhancing non-linear optical interactions \cite{li2024inverse}, decoupling modes in waveguides \cite{Qiao2025, Zalogina2025}, optical computing \cite{Cordaro2023equation}, spatial bandpass filters \cite{Clark2025} and image processing in the mid-infrared \cite{Pearson2025}. The algorithm can also be produce metasurfaces robust to fabrication defects \cite{phan2019high, wang2019robust}. Topology optimisation, however, has not been applied to developing devices for the imaging of phase.

Here, we design and demonstrate a non-local, free-form dielectric metasurface that switches from a brightfield to a phase-imaging modality by changing the illuminating polarisation. When illuminated with 886 nm $x$-polarised light, the metasurface performs the identity operation on the field, corresponding to conventional brightfield imaging. Under the orthogonal $y$-polarised illumination, the metasurface performs a first-order derivative with respect to $x$ permitting visualisation of the phase gradient introduced into the incident field by the sample. Our metasurface performs these operations in the object plane, directly filtering the spatial frequency information of the image in an ultra-compact form-factor. We employ topology optimisation to design a metasurface with both high performance and robustness to fabrication defects. In addition, unlike previous work featuring dielectric metasurfaces, our metasurface performs computational operations at normal incidence. Avoiding tilting the metasurface or sample, or the use of off-normal illumination, requires an asymmetry in the transmission about normal incidence, which is here generated through diffraction. We choose the period of the metasurface so that higher order diffracted modes are trapped by total reflection inside the substrate which generates an asymmetric transmission in the zeroth order. The zeroth order emerges into free space and can be collected by the imaging system. We fabricate the metasurface and demonstrate tunable image processing and the computational ability of our metasurface to visualise synthetic phase test targets as well as biological cells and tissue samples. 

\begin{figure}[H]
    \centering
    \includegraphics[width=\linewidth]{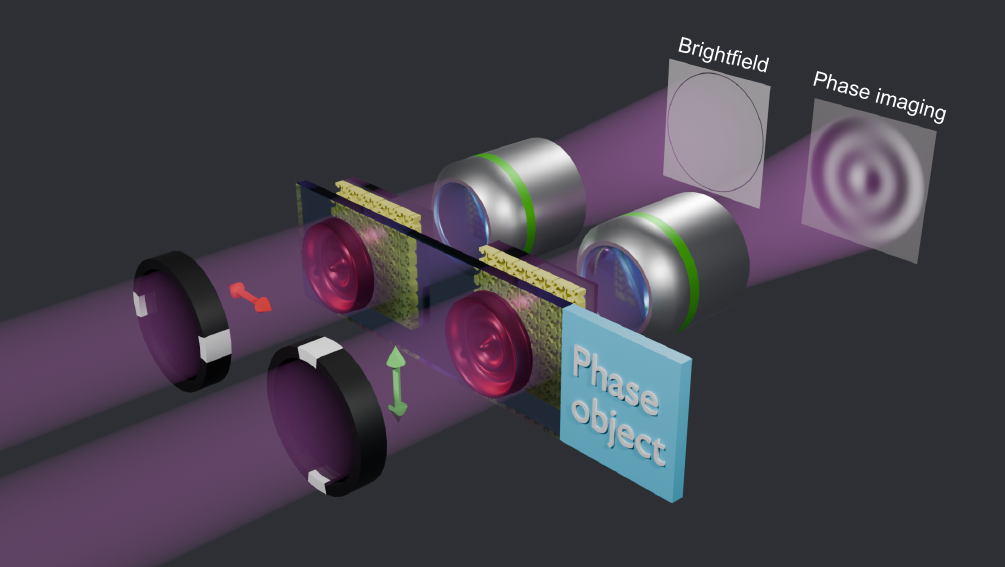}
    \caption{Schematic of the switchable image processing. By simply rotating a polariser, the metasurface enables switchable brightfield and phase contrast imaging.}
    \label{fig:schematic}
\end{figure}

\section{Results}

\subsection{Spatial filtering}

Using the angular spectrum decomposition of an optical field \cite{Goodman1996}, we can write an arbitrary monochromatic, spatially coherent optical field as a superposition of plane waves. These plane waves are characterised by their transverse wavevector components $(k_x,k_y)$ referred to as their spatial frequencies. Rapidly varying details in an image, such as edges, are encoded by high spatial frequencies while slowly varying details are encoded in low spatial frequencies. By selectively modifying the amplitude and phase of the plane waves making up an optical field, we perform various operations on the resulting image. The momentum (spatial frequency) response of an optical system is described by the optical transfer function (OTF). In a scalar system, if the optical system is linear and spatially invariant and the illumination is spatially and temporally coherent, then the output field $U(x, y)$ is given by:
\begin{equation}
    U(x, y) = \mathcal{F}^{-1} \left[ H(k_x, k_y) \mathcal{F} \left[ U_0(x, y) \right] \right],
\end{equation}
where $\mathcal{F}$ and $\mathcal{F}^{-1}$ denote the two-dimensional spatial Fourier and inverse Fourier transforms respectively, $H(k_x, k_y)$ is the OTF and $U_0(x, y)$ is the input field. By engineering the OTF of an optical system, images can be modified in various ways. For example blocking low, while transmitting high, spatial frequencies produces edge enhancement. If the optical system is polarisation sensitive then we can describe the propagation of two orthogonal polarisations through the optical system in terms of the co- and cross-polarised OTFs,

\begin{equation}
    \begin{bmatrix}
        U_s (x, y) \\
        U_p (x, y)
    \end{bmatrix}
    = \mathcal{F}^{-1}\left[
    \begin{bmatrix}
        H_{ss}(k_x, k_y) & H_{ps}(k_x, k_y) \\
        H_{sp}(k_x, k_y) & H_{pp}(k_x, k_y) 
    \end{bmatrix} 
    \mathcal{F} \left(
    \begin{bmatrix}
        U_{0s} (x, y) \\
        U_{0p} (x, y)
    \end{bmatrix}
    \right)
    \right]
\end{equation} 
expressing the response in terms of an $s$ and $p$ polarisation basis. In the above expressions, the OTFs for the co-polarised light, $H_{ss}(k_x, k_y)$ and $H_{pp}(k_x, k_y)$ lie on the main diagonal of the tensor and the off diagonal components, $H_{ps}(k_x, k_y)$ and $H_{sp}(k_x, k_y)$ represent the cross-polarised OTFs for incident $s$ polarised light converted into $p$ polarised light and vice versa. 

To design a metasurface with a specific OTF, we need to produce a structure with a transmission sensitive to the transverse momentum of the light. There is a one-to-one correspondence between spatial frequency and the angle the corresponding plane wave travels with respect to the optical axis, taken here to be the $z$-axis. The spatial frequencies, $k_x$ and $k_y$, can then be written in terms of angular coordinates via,
\begin{align}
    k_x =& k_0\sin{\left( \theta \right)} \cos{\left( \varphi \right)}\\
    k_y =& k_0\sin{\left( \theta \right)} \sin{\left( \varphi \right)},
\end{align}
where $\theta$ is the elevation angle with respect to the optical axis, $\varphi$ is the azimuthal angle, $k_0=2\pi/\lambda$ is the wave-number and $\lambda$ the wavelength in free space. Due to the correspondence between spatial frequency and angle, engineering the OTF corresponds to tailoring the angular transmission of the metasurface.  

Consider the following (scalar) OTF, $H(k_x, k_y)=\alpha k_x + \beta$, where $\alpha$ and $\beta$ are real constants. In an optical system with no gain, $|H(k_x, k_y)|\leq 1$. In the case of imaging a pure phase object, the transmitted field has a constant amplitude and is of the form, $U_0(x, y) =A e^{i \phi(x, y)}$, where $A$ is a constant and $\phi (x, y)$ is the phase introduced to the optical field by the transparent object. In this case, the field and intensity that emerge from the metasurface are, respectively, given by, 
\begin{align}
    U(x, y) &= U_0(x,y) \left( \beta + \alpha \frac{\partial \phi(x,y)}{\partial x} \right) \\
    I(x, y) &= A^2 \left( \alpha^2 \left(\frac{\partial \phi(x, y)}{\partial x}\right)^2 + \beta^2 + 2\alpha \beta \frac{\partial \phi(x, y)}{\partial x}\right).
    \label{eq:linear_otf_effect}
\end{align}
Using this linear form of the transfer function, we convert phase gradient information along the $x$-direction into an intensity modulation which can be detected. Furthermore, a non-zero $\beta$ gives information about the sign of the phase gradients providing rich information about the specific sample and producing pseudo-3D images, similar to those obtained from DIC. If desired, this OTF can also be used to quantify the first-order derivative, permitting all-optical analogue optical computation and providing a tool to visualise transparent specimens. 
 
Non-local periodic metasurfaces are a perfect candidate for implementing these momentum sensitive operators since the angular response is uniform across the surface and structuring the sub-wavelength unit cells gives rich control over the optical properties. This approach permits object plane image processing, with a large degree of freedom over where the metasurface is placed in an optical system.

\subsection{Metasurface Design}

The metasurface is designed with a topology optimisation algorithm that considers robustness to dilation and erosion fabrication defects, inspired by \cite{Kim2024}. The unit cell is decomposed into pixels where we optimise the material density at each pixel directly. The material density, after undergoing several filtering steps to promote binarisation and enforce minimum feature sizes \cite{Wang2010projectionmethods}, represents the permittivity distribution of the unit cell, where a value of zero corresponds to air and a value of one corresponds to amorphous silicon (optical constants \cite{Pierce1972}). Rigorous coupled wave analysis (RCWA) (as implemented in TORCWA \cite{torcwa}) is used to evaluate the OTF and we calculate the loss function, taken here to be the root mean square error (RMSE) between the current OTF and the desired OTF. We use the automatic differentiation engine built into PyTorch to calculate the gradient of the loss function with respect to the material density at each pixel location which is then fed into a gradient based optimisation algorithm to update the material density. Robustness considerations are made by performing a weighted average of the cost function over the dilated, regular and eroded designs. The eroded and dilated designs are generated in a three step process. Edge detection is performed on the material density, which is then blurred and then added or subtracted from the density to form the dilated and eroded variant respectively. The amount of blurring applied to the edges controls the amount of erosion or dilation. The full pattern generation and optimisation procedure is explained in the supplementary information.

We design the metasurface with a polarisation switchable OTF. For $x$-polarised light, the device produces a constant OTF corresponding to brightfield imaging while for $y$-polarised light the OTF is linear in $k_x$ which is the phase imaging modality. The range of angles over which we optimise the OTF defines the numerical aperture (NA) and hence the spatial resolution of our imaging system, The numerical aperture is given by, $NA = \sin{\left( \theta_{\mathrm{max}}\right)}$, where $\theta_{\mathrm{max}}$ is the maximum angle we consider. The choice of NA is informed by several factors. It defines the resolution of our imaging system and is determined by the sizes of the samples we wish to image. Secondly, there is a trade-off between the image contrast and NA. When we optimise for a linear OTF, as the NA increases, the gradient of the linear OTF must decrease, ($\alpha$ in equation \ref{eq:linear_otf_effect}), and the visibility of the processed image decreases. We choose to optimise over 5 evenly spaced angles from $-3 \degree$ to $+3 \degree$ which corresponds to an NA of 0.05, targeting the two OTFs given by 
\begin{align}
    H_{xx}(k_x) &= \gamma \label{eq:otf_xx}\\
    H_{yy}(k_x) &= \alpha k_x/k_0 + \beta \label{eq:otf_yy},
\end{align}
where we choose $\gamma=0.8$, $\beta=0.5$ and $\alpha=0.5/\mathrm{NA}$.

\begin{figure}[ht]
    \centering
    \includegraphics[width=\linewidth]{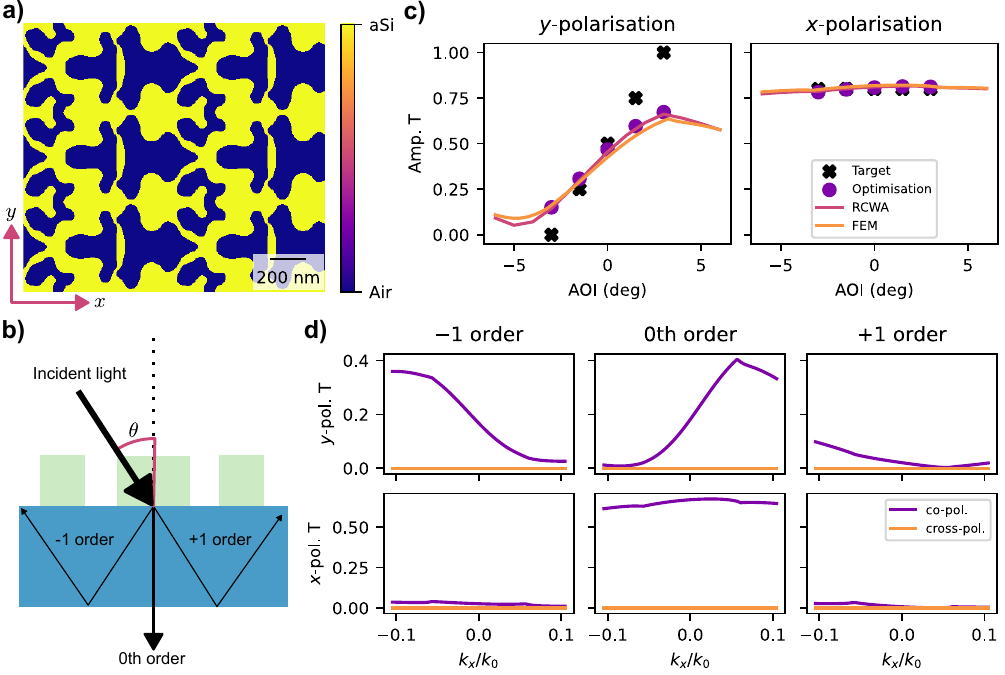}
    \caption{Design and simulation of the switchable image processing metasurface. \textbf{a)} Final design of the metasurface, \textbf{b)} schematic of the diffracted orders that are use to generate the asymmetric OTF. \textbf{c)} 1D OTFs for $x$ and $y$-polarised light simulated with RCWA and FEM compared to the target OTF. \textbf{d)} Diffraction efficiency as a function of spatial frequency, simulated using the FEM}
    \label{fig:des_sim}
\end{figure}

We optimise the OTF for 900 nm illumination. At this wavelength, silicon acts as a low-loss dielectric, yet silicon based photodetectors can still be utilised for the imaging experiments. Our metasurface consists of a patterned 300 nm thick layer of amorphous silicon on a quartz substrate (optical constants \cite{Malitson1965}). Light is incident from the air side and the transmission is measured in the substrate. Our target OTF requires an asymmetric response about normal incidence. One mechanism to generate asymmetry is through the use of chiral metasurfaces \cite{Wesemann2022asym}, whereas here we use diffraction effects. We choose a period, $850$ \um~$\times$ $500$ \um, permitting diffraction into the $\pm 1$ diffracted orders which propagate along the $k_x$ direction within the quartz substrate. Asymmetric, angle dependent diffraction occurs into these higher orders which are totally internally reflected inside the substrate, a schematic of which is shown in figure \ref{fig:des_sim}b. Diffraction removes energy from the zeroth order, resulting in an OTF approximately linear from $-3$ to $+3$ degrees. We want the OTF to be independent of $k_y$, so we chose a sub-wavelength period in the $y$-direction to prevent diffraction and force symmetry about the $x$-axis to ensure that the OTF in the $k_y$ direction is symmetric. During the optimisation we use $33 \times 21$ orders in the RCWA expansion. The resulting metasurface design and its optical properties are presented in figure \ref{fig:des_sim}. 

The optimisation procedure produces a metasurface with an OTF with the anticipated imaging capabilities, as shown in figure \ref{fig:des_sim}c. For $x$-polarised light, the OTF of the metasurface is flat with a RMSE from the target of 0.012, whereas for $y$-polarised illumination the OTF is approximately linear as desired. There is, however, a loss of contrast relative to the target and the RMSE is 0.178. Contrast refers to the difference between the maximum and minimum transmission in the OTF and is directly related to the strength of the phase gradients in imaging experiments. Our optimisation process, produces a structure with a contrast of 0.52, still suitable for image processing applications. Although there are discrepancies between the OTF and the target, the design is quite robust to small fabrication defects improving fabrication outcomes (see supplementary information). The optical response is validated with a subsequent RCWA simulation with $41 \times 31$ Fourier orders and a finite element method (FEM) simulation (COMSOL Multiphysics), which reproduce the optimised behaviour. Furthermore, we use the FEM to calculate the diffraction efficiency as a function of spatial frequency (figure \ref{fig:des_sim}d). The asymmetry is primarily driven by diffraction into the $-1$ order which propagates in the $-k_x$ direction. From ray tracing simulations (see supplementary information), the diffracted orders undergo total reflection inside the substrate which acts as a waveguide. At larger angles of incidence, the diffracted orders can escape from the substrate, but travel at large angles and are not captured by the NA of the imaging system. The asymmetric diffraction and use of the substrate to carry away the higher orders leads to the required asymmetry in the zeroth order. This diffraction response disappears for $x$-polarised illumination, enabling the polarisation switchable behaviour. The full 2D OTF of the metasurface is only weakly dependent on $k_y$ and the phase response is reasonably constant for both polarisations within the processing NA (see supplementary information).

\subsection{Metasurface characterisation}

The metasurface is fabricated with a top-down approach. Amorphous silicon is deposited with plasma-enhanced chemical vapour deposition, then electron beam lithography with PMMA as resist is performed to define the pattern on the surface. A 20 nm layer of Al$_2$O$_3$ is deposited using electron beam evaporation to serve as a hard mask, which is lifted off and the pattern etched into the silicon using a pseudo-Bosch process. Details provided in the Methods section.

\begin{figure}[ht]
    \centering
    \includegraphics[width=\linewidth]{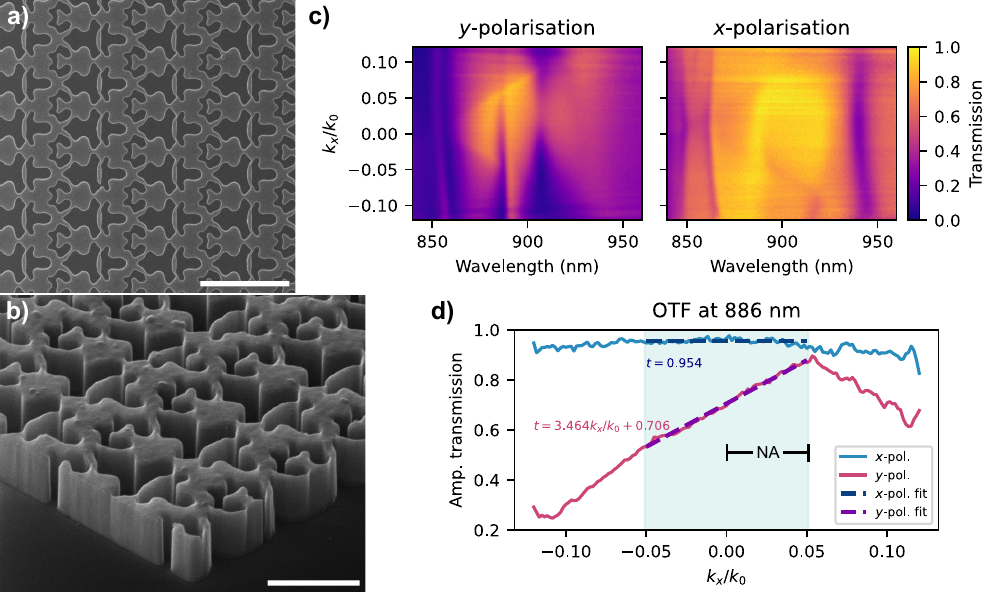}
    \caption{Fabricated topology optimised structure. \textbf{a)} Helium-ion microscope (HIM) image taken at normal incidence, scalebar is 1 \um, \textbf{b)} HIM image taken with sample tilted by 55\degree, scalebar is 500 nm. \textbf{c)}, Transmission spectrum of the device as a function of normalised spatial frequency, \textbf{d)} OTF of the metasurface at 886 nm illumination.}
    \label{fig:fab_char}
\end{figure}

The fabricated structure is imaged using a Helium ion microscope (HIM), with micrographs given in figure \ref{fig:fab_char}a and b. Looking at the tilted HIM image, we can see there is some surface roughness which is a part of the Al$_2$O$_3$ etch mask layer. We choose not to remove the etch mask layer since it is both thin and transparent. In addition, simulations (supplementary information) show that this layer has minimal effect on the optical properties of our device. 

The 1D OTF as a function of wavelength (figure \ref{fig:fab_char}c) is obtained by imaging a slice of the back focal plane with an imaging spectrograph (Shamrock 303i spectrograph equipped with an Andor iDus CCD). There are two regions where the OTF exhibits an asymmetry about normal incidence, at 886 nm and at 900 nm. At 886 nm, the resulting OTF is approximately linear across the NA of 0.05 as desired and at 900 nm, the OTF tapers off for positive spatial frequencies and is no longer linear. While this behaviour is useful for imaging transparent samples qualitatively, the ability of this metasurface to perform computational operations at this wavelength is limited. Therefore, we perform the imaging experiments at 886 nm, and the one-dimensional OTF at this wavelength is given in figure \ref{fig:fab_char}d. The full 2D modulation transfer function (MTF) at 886 nm is measured by imaging the entire back focal plane and is presented in the supplementary information. As simulated, the 2D OTF varies along the $k_x$ direction and there is little variation along $k_y$. For $x$-polarised light, the OTF is constant across an NA of 0.1 and for $y$-polarised light, the linear region of the OTF exists over an NA of 0.05. At larger spatial frequencies, $k_x/k_0 > 0.05$, the OTF begins to decrease, whereas for $-0.1 < k_x/k_0 < -0.05$, the OTF continues the linear trend. We fit a constant to the $x$-polarised OTF and do a linear fit to the $y$-polarised OTF over the 0.05 NA, the equations of which are given below. 
\begin{align}
    t = (0.954\pm0.006)&,\quad x\mathrm{- polarisation} \label{eq:OTF_x_m}\\
    t = (3.464 \pm 0.150) k_x/k_0 + (0.706 \pm 0.005) &,\quad y\mathrm{- polarisation} \label{eq:OTF_y_m}
\end{align}
where $t$ represents amplitude transmission and $k_x/k_0$ is the normalised spatial frequency. Fitting is performed with the Numpy polyfit function and errors represent 5 standard deviations. These coefficients can be used to quantitatively recover the phase gradient from the phase contrast images of transparent samples. Simulations show that the phase response of our metasurface is constant within our imaging NA, so we assume the metasurface OTF is real when we perform recovery of the phase gradients.

Although the characteristics of the OTF are the same as the design properties, the values of the coefficients are different. In particular, the transmission for both $x$ and $y$-polarised light at normal incidence is greater than in simulations. For phase imaging, the most important aspects of the OTFs are the slope and the NA, which agree well with the design. For other image processing operations, such as edge detection, transmission at normal incidence is the most important factor. Furthermore, the operational wavelength has blueshifted slightly. The fabricated geometry is slightly dilated compared to the design geometry, however, this dilation is non-uniform (see supplementary information).

\subsection{Imaging of phase test targets}

\begin{figure}[H]
    \centering
    \includegraphics[width=\linewidth]{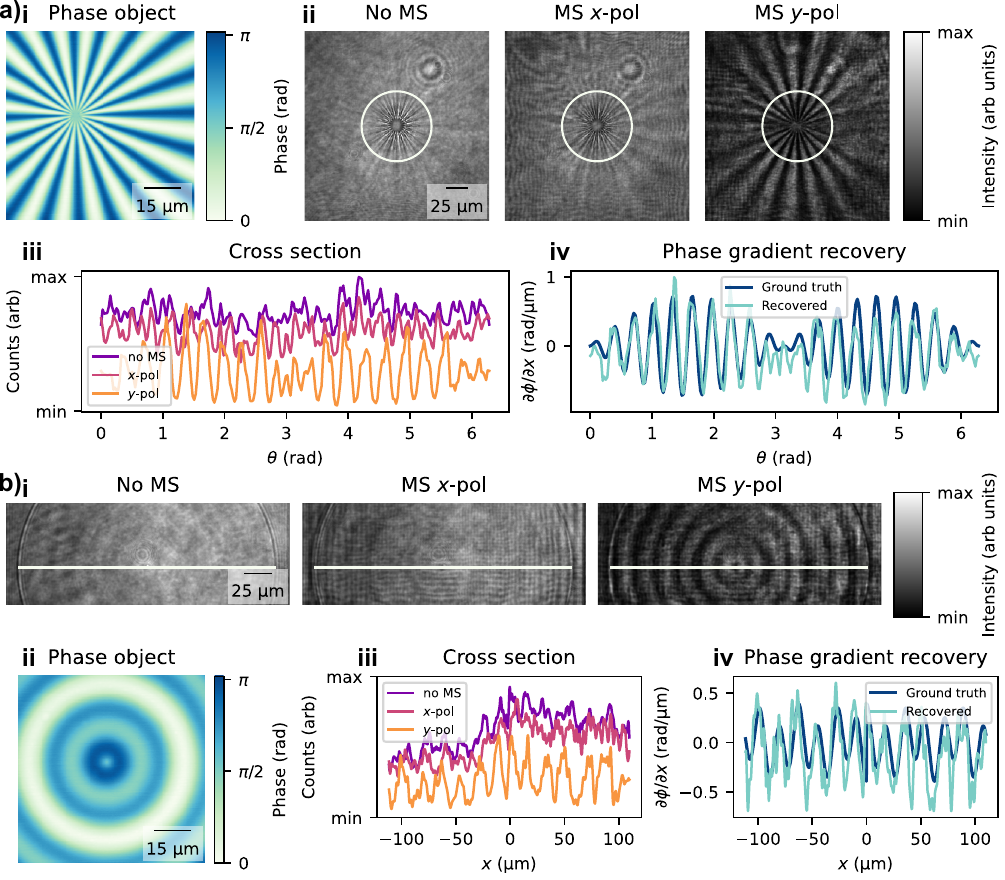}
    \caption{Imaging experiments performed on fabricated phase test targets. \textbf{a)} Images of a phase spoke target with an excursion of $\pi$ rad at 886 nm. \textbf{i} AFM image of the phase object converted into phase, \textbf{ii} image of the target without metasurface and with metasurface, \textbf{iii} cross section along the circle outlined in \textbf{ii}, \textbf{iv} recovered phase gradient around the cross section. \textbf{b)}, Imaging on a phase ``bullseye" structure. \textbf{i}, images with and without the metasurface, \textbf{ii} AFM image of the phase object converted into phase, \textbf{iii} cross section of the intensity along the line in \textbf{i}, \textbf{iv} recovery of the phase gradient along the cross-section line.}
    \label{fig:lith_imaging}
\end{figure}

We demonstrate switchable image processing on fabricated phase targets. Using greyscale lithography we fabricate phase test targets designed to have a maximum phase excursion of approximately $\pi$ radians at 886 nm (fabrication details in Methods). Since our resist is transparent with a well characterised refractive index, we can control the phase by generating specific topographies. We fabricate two phase objects, a spoke target (figure \ref{fig:lith_imaging}a i) and a ``bullseye" structure (figure \ref{fig:lith_imaging}b ii). The equations defining these objects, topography and refractive index data are available in the supplementary information. The fabricated phase objects have profiles that are in excellent agreement with the design, although the height is slightly less than anticipated (see supplementary information). 
 
Image processing experiments are performed with the metasurface located in the object plane in a custom microscope setup (see supplementary information). A polarised collimated beam is incident on the phase object which is brought to sharp focus. The light then passes through the metasurface which is positioned as close as possible to the phase object to reduce diffraction artefacts from its edges. Imaging is performed with a 20$\times$ long working distance objective and a tube lens forms the image on the camera. To align the metasurface normal to the beam, we adjust the tilt of the metasurface such that the reflected light is coincident with the illumination, which is performed at 690 nm to permit visual inspection.

From the images obtained with the two polarisation states, we can recover the gradient in the phase along the $x$-direction. When imaging a phase object, for $y$-polarised light, the intensity we measure depends on the phase gradients and a background term, see equation \ref{eq:linear_otf_effect}. For $x$-polarised light, we perform an identity operation permitting a measurement of the background without removing the metasurface. We can quantitatively recover the phase gradients by solving the quadratic equation for the phase derivative.
\begin{equation}
    \frac{\partial \phi}{\partial x} = \frac{1}{\alpha} \left( \sqrt{\frac{\gamma^2 I_y - \beta^2 I_x}{I_x} + \beta^2} - \beta\right),
\end{equation}
where, $I_x$ is the $x$-polarised image and $I_y$ is the $y$-polarised image. The coefficients of the transfer functions, $\alpha, \beta, \gamma$ are taken from the experimental measurements of the OTF (equations \ref{eq:OTF_x_m} and \ref{eq:OTF_y_m}). Recovering the phase gradient in this manner requires near perfect alignment of the metasurface. Tilting the metasurface by $0.5\degree$ results in a change in $\beta$ of $0.09$. To remove this tilt, we assume that the average phase gradient in the samples is zero and we adjust the parameter $\beta$ until the average phase gradient is less than 0.01 rad/\um. Since we align the metasurface to be perpendicular to the beam, the amount of tilt is small and this fitting process converges quickly. Details about this optimisation procedure are in the supplementary information. Tilt removal is a mature idea and other more sophisticated algorithms, such as three point levelling  could also be used. A table containing the fitting parameters and implied tilts for the imaging experiments is provided in the supplementary information.

Images of the fabricated test objects are given in figure \ref{fig:lith_imaging}a ii and \ref{fig:lith_imaging}b i. We can see that without the metasurface, slowly varying features in the phase object are invisible. Rapidly varying details, such as the centre of the spoke target and the edge of the ``bullseye" structure, are visible due to the limited NA of the imaging system which blocks high spatial frequencies producing imaging artefacts. When illuminating the metasurface with $x$-polarised light the image is almost identical to that obtained in the absence of the device. There are, however, some diffraction artefacts that appear in the image due to both small defects and from scattering by the edges of the metasurface. When illuminating with $y$-polarised light there is horizontal contrast generated that depends on the phase gradients in the object. The resulting images have a pseudo-3D appearance, which is more apparent with the ``bullseye" phase object. Details due to variations in the phase in the $y$-direction disappear in both images, since only the gradients along the $x$-axis generate contrast. 

We quantitatively recover the phase gradient from the spoke images and plot a cross-section of the phase gradient along the circle outlined in figure \ref{fig:lith_imaging}a ii starting at the right-most point of the circle and going anti-clockwise. We parametrise the position on this arc by $\theta$, representing the angular distance around the path from the starting point. We can see that the cross-sections for $x$-polarised light and without the metasurface are very similar in profile but with a small offset, due to the $x$-polarised OTF not producing 100\% transmission. There is a periodic signal in the $y$-polarised cross-section which is a maximum at $\theta = \pi/2$ and $\theta = 3\pi/2$ corresponding to the locations of maximum gradient along the $x$-direction in the phase object. The phase gradient is recovered and presented in panel iv of figure \ref{fig:lith_imaging}a. There is a discrepancy between the behaviour of the ground truth and recovered gradients at $\theta = \pi$, where there is a small defect in the metasurface. Around this circle, the phase gradient recovery has a RMSE of 0.0548 rad \um$^{-1}$.

The ``bullseye" image has non-uniform illumination, where the left side is dimmer than the right side. This non-uniformity has no effect on the recovery of the phase gradients from the sample, along the cross-section line, the RMSE of the recovery is 0.0617 rad \um$^{-1}$. Again, the largest error in the recovery is due to noise from diffractive effects. In both cases, we compare the phase gradient recovered with the analytical expression obtained by differentiating the equation defining the phase objects, which are adjusted to accommodate the amplitude of the fabricated samples. We can successfully quantitatively extract phase gradient information using our metasurface. The presence of noise, however, negatively impacts our recovery ability. 

\subsection{Imaging of biological samples}

To test the imaging performance of our metasurface on biological samples, we perform imaging of HeLa cells and a sample of human ovarian cancer tissue. Biological samples can have relatively weak phase excursions and contain small features, to prevent these features being hidden in noise, we reduce the diffraction artefacts from the metasurface by averaging over 10 images in which the metasurface is moved by a small offset. This softens the influence of the metasurface edges, reducing noise, while the sample remains fixed and sharply in focus. We present these averaged images in the main text and images without averaging are provided in the supplementary material. Since our metasurface is non-local and the angular response is identical across the surface, there is no impact from averaging over the metasurface on the operations on the sample.

\begin{figure}[H]
    \centering
    \includegraphics[width=\linewidth]{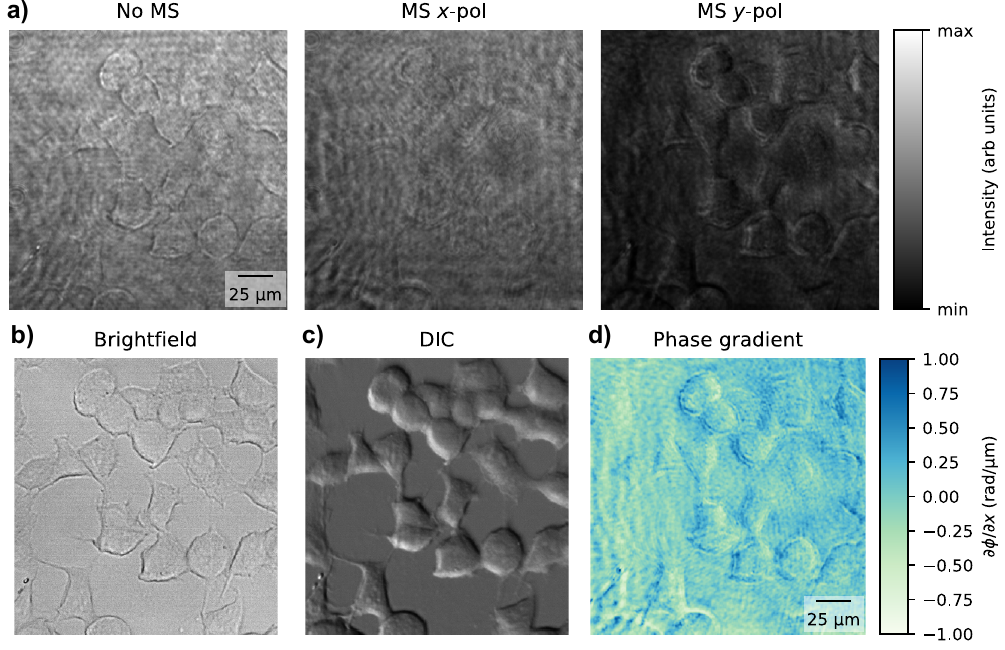}
    \caption{Imaging of HeLa cells with the metasurface placed in the object plane. \textbf{a)} Image of the cells without and with the metasurface, \textbf{b)} and \textbf{c)} brightfield and differential interference contrast images of the cells taken with Olympus BX60 microscope. \textbf{d)} Recovered phase gradient along the $x$-direction using the metasurface.}
    \label{fig:hela_ims}
\end{figure}

The results of imaging experiments performed on a sample of unstained HeLa cells are presented in figure \ref{fig:hela_ims}. Without the metasurface present, these cells are very difficult to detect, with only some contrast around their edges. With the metasurface added, using $x$-polarised light produces a similar image and we cannot garner information about the cell morphology. When switching to $y$-polarised light, the location and morphology of the cells becomes much clearer, with the left side of features becoming bright and their right side dark. Further noise and speckle is removed during the phase gradient recovery process, with \ref{fig:hela_ims}d demonstrating a very clear image of the cells. The image generated by the metasurface under $y$-polarised illumination is in good agreement with that obtained from a microscope set up to perform differential interference contrast (DIC) microscopy, figure \ref{fig:hela_ims}c. DIC generates contrast along a $45\degree$ diagonal, rather than along the $x$-axis as is the case for the metasurface. Furthermore, DIC has superior spatial resolution with an NA of 0.4, compared to the 0.05 of the metasurface.

\begin{figure}[H]
    \centering
    \includegraphics[width=\linewidth]{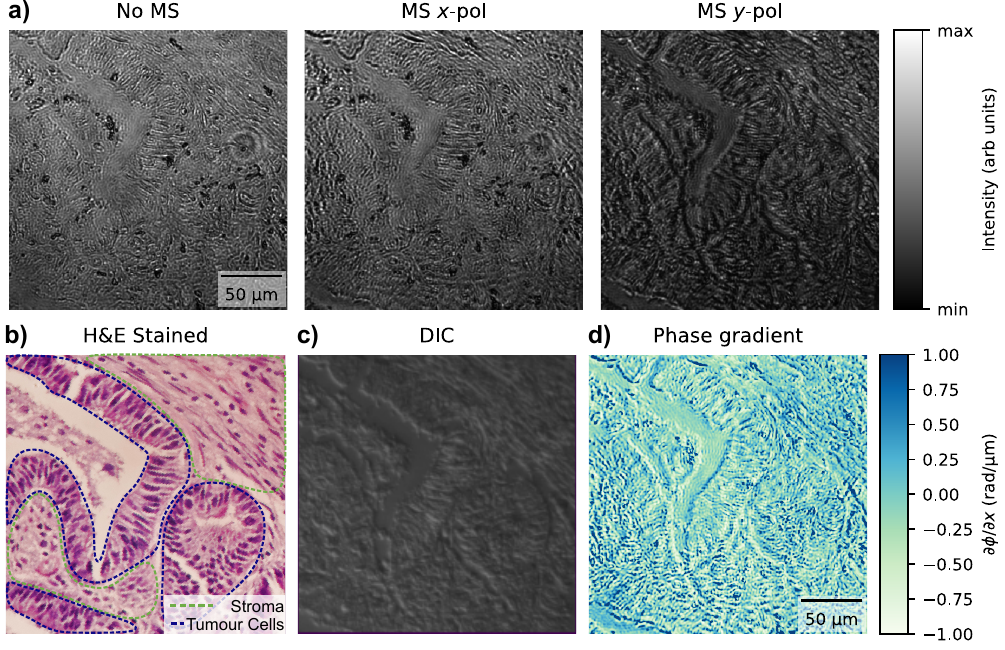}
    \caption{Imaging of an ovarian cancer sample with the metasurface placed in the object plane. \textbf{a)} image of the sample without and with the metasurface, \textbf{b)} H\&E stained sample cut from an adjacent section, \textbf{c)} differential interference contrast image taken with Olympus BX60 microscope. \textbf{d)} Recovered phase gradient along the $x$-direction using the metasurface.}
    \label{fig:tissue_ims}
\end{figure}

We also obtained images of an unstained sample of human mucinous ovarian adenocarcinoma, images of which are shown in figure \ref{fig:tissue_ims}. Unlike the HeLa cells, in the brightfield images, both without the metasurface present and metasurface with $x$-polarised light, some weakly absorbing features can be visualised. Again, when the incident polarisation is rotated to be vertical, we can see a pseudo-3D image of the sample and we can begin to observe the textures within. In particular, the gaps within the tissue, the intra-cystic spaces, can be differentiated from the surrounding cells quite easily. Furthermore we can begin to see individual cells within the tissue section. The circular arrangement of tumour cells on the right side of the image can be distinguished using the metasurface with $y$-polarised light. Without the metasurface present, or under $x$-polarised illumination, details within this region are difficult to identify. Using the metasurface we gain a greater understanding of the orientation and morphology of the cells within the tissue without using labels in an ultra-compact form factor. The recovered phase gradient image and the metasurface $y$-polarised image are in good agreement with the DIC image.

\section{Discussion}
We have demonstrated a metasurface produced using inverse design that can simultaneously perform two computational operations on images directly in the object plane simply by rotating the illumination polarisation and without adding extra optical components. We can switch between brightfield imaging, an identity operation, and phase imaging, via a first order derivative. The asymmetry about normal incidence required for this operation is driven by diffraction. We have demonstrated the switchable imaging and computational ability of our metasurface on synthetic phase objects and biological samples.

There are small differences between the simulations and fabricated metasurface. During the optimisation, we consider only uniform edge dilation and erosion, variation from the optimised performance can occur due to non-uniform design defects and small mismatches between the optical properties of the silicon films used in simulations and those deposited. Robustness algorithms that model realistic fabrication outcomes \cite{Piggott2020, Schubert2022} can be employed to reduce the gap between design and fabrication and ensure designs are compatible with scalable fabrications methods such as EUV photolithography.

Topology optimisation gives us precise control over the numerical aperture of the image processing operations. Due to the trade-off between NA and imaging contrast, topology optimisation can be used to create imaging metasurfaces tailored to specific sizes of samples. Tissue samples generally have larger features, a smaller metasurface NA can be used for the imaging to maximise contrast compared to cell imaging. Our approach using diffraction to generate a linear OTF provides a simple mechanism to generate an asymmetric angular response which can be extended to larger or smaller NAs and other transfer functions where an asymmetric sensitivity to angle of incidence is required. Furthermore, topology optimisation could be used to tailor a transfer function to detect specific features in a sample for object recognition tasks. In addition, this approach can be extended to operations in the image plane, where the metasurface is adjacent to the camera sensor to design computational camera systems. In general, image plane image processing would require a metasurface with a very small processing NA consistent with the resolution associated with the pixel spacing. An image plane approach would forgo the requirement to tailor the NA to the object being imaged, since an appropriate objective in the imaging system can be used to resize the object to a suitable size. Future work can also focus on designing broadband non-local filters. A metasurface that can perform non-local spatial filtering over a range of wavelengths can be used with an LED illumination source, enabling an ultra-low cost imaging system. Our approach requires spatially coherent light, which can be generated by spatially filtering the LED source with a pinhole in a $4f$ system or potentially with another angle-sensitive metasurface to maintain the compact size of the system.

With the non-local object plane imaging approach, we are afforded a great deal of freedom in positioning the metasurface within the optical system. In Fourier optics, as the light propagates, the plane waves making up the angular spectrum of the field acquire a spatial frequency dependant phase term. The object plane approach outlined here modifies the amplitudes of these plane waves. As long as the phase transfer function is constant or a weak function of spatial frequency, such as our design, defocus does not have an impact on the image processing. Defocus can however introduce diffraction artefacts resulting from the finite size of the metasurface. However, if the metasurface is sufficiently large, then the diffraction effects can be sufficiently minimised across a smaller field of view, see the supplementary information.

The errors and artefacts in our phase gradient recovery are dominated by noise which arise primarily from small defects in the metasurface. Although small, since the metasurface is not in focus, diffraction amplifies the effect of these defects. Taking several images with the metasurface in slightly different locations permits averaging this noise, but requires several images to be taken. Improving the fabrication process would result in a smoother isotropic metasurface avoiding this problem.

Furthermore, in some cases with a single measurement of the gradient in one-direction it is possible to recover the full phase distribution \cite{MatasDiMartino2013} with reasonable results. This approach requires knowledge about the particular samples being imaged and assumes there is little information located on the $k_y$ axis in spatial frequency space. For the two artificial samples presented in this work, this assumption is not valid. To perform quantitative phase imaging on arbitrary samples, the derivative along both $x$ and $y$ needs to be computed. Future work could focus on designing a metasurface to recover gradients along both the $x$ and $y$ directions by changing the polarisation to enable quantitative phase imaging.

\section{Conclusion}

We have demonstrated a non-local freeform metasurface that is engineered to switch between a brightfield, identity operation and a phase imaging, first-order derivative operation simply by changing the polarisation. Topology optimisation is a powerful approach to designing metasurfaces for image processing and the use of non free-space propagating diffracted orders to generate asymmetry can be tailored to imaging different types of samples in both the object plane and image plane and implementing other asymmetric transfer functions such as those performing odd-order differentiation. We envisage such metasurfaces will enable ultra-compact, low-cost imaging systems and enable energy-free image processing operations.

\section{Methods}

\subsection{Metasurface fabrication}

The metasurface is produced in a top-down fabrication process. First, 300 nm of amorphous silicon is deposited on a diced quartz wafer with PECVD (Oxford PlasmaPro 100). We spincoat 200 nm of PMMA which is subsequently baked at 180 $\degree$C for 5 minutes. After cooling to room temperature, we spincoat a water-soluble conductive coating (Dischem Discharge H2Ox2) and a permanent marker is used on the backside of the substrate to render it opaque. EBL is performed on a Vistec EBPG5000plus with a beam step size of 2 nm, a spot size of 3 nm and a dose of 700 \textmu C cm$^{-2}$. Each metasurface measures $300\times300$ \um$^2$. The PMMA is developed with a 1:3 solution of MIBK:ethanol for 60~s, then rinsed in isopropanol for 15 s and DI water for a further 15 s. 

We then evaporate a 20 nm layer of Al$_2$O$_3$ to serve as a hard mark using electron beam evaporation (Intlvac Nanochrome II). To perform liftoff, the sample is placed into acetone for 24 hours and then ultrasonicated for 20 s, rinsed with IPA and then DI water. The sample is then etched for 30 s in an Oxford Instruments PlasmaPro 100 Estrelas using a Pseudo-Bosch process (20:21 ratio of SF$_6$ to C$_4$F$_8$).

The resulting metasurface is imaged using a helium-ion microscope, Zeiss ORION NanoFab.

\subsection{Phase test target fabrication}

Phase test targets are fabricated using greyscale lithography in resist AZ5214E using a POLOS NanoWriter Advanced. We spincoat 1.4 \um~of resist onto a glass coverslip and exposure is performed with a maximum dose of 45 mJ cm$^{-2}$, a spotsize of 300 nm, step size of 150 nm and at a wavelength of 405 nm. The resulting sample is developed in AZ MIF 726 for 60 s and rinsed with DI water. The resulting topography is measured with an AFM (MFP-3D).

The phase $\phi$ of the sample is calculated using the expression
\begin{equation}
    \phi (x, y) = \frac{2\pi \Delta n}{\lambda} T(x, y) \label{eq:phase},
\end{equation}
where $T(x,y)$ is the thickness of the sample at $(x,y)$, $\lambda$ is the operating free-space wavelength and $\Delta n$ is the difference in refractive index between air and the photoresist. The refractive index of the photoresist is given in the supplementary materials. 

\subsection{Cell sample preparation}

Coverslips with dimensions 22 $\times$ 22 mm were placed in a 6-well dish. They were then coated with 0.1 mg/mL Poly-D-Lysine (PDL) for 30 minutes at room temperature. HeLa cells were grown in DMEM (Lonza) supplemented with 10\% bovine growth serum (Gibco), 1 $\times$ Pen-Strep (Lonza) at 37°C in 5\% CO$_2$. Cells were fixed with 4\% paraformaldehyde for 15 minutes at room temperature. After fixation, coverslips were transferred onto glass slides for imaging.

\subsection{Tissue preparation}

A thin section (4 micron) was cut from a formalin-fixed and paraffin embedded tissue microarray containing human mucinous ovarian adenocarcinoma samples and mounted on Superfrost Plus adhesive slides (Thermo Scientific). An adjacent serial section was stained by haematoxylin and eosin (H\&E) using standard procedures.

To prepare the tissue sections sample for imaging experiments, deparaffinisation and rehydration was performed through a series of xylene, 100\% ethanol and RO water washes before being dried at room temperature overnight. To reduce scattering from features within the tissue sample, we place a drop of water on the fixed tissue sample and attach a coverslip prior to imaging.

\section*{Ethics Approval}
Analysis of ovarian cancer tissues was approved by the Peter MacCallum Cancer Centre HREC, project \#20/34. Patients provided informed consent for the use of their tissue for research.

\section*{Supplementary Materials}

The supplementary information contains details about the topology optimisation algorithm, additional metasurface simulations, additional experimental results, and the optical setups. 

\section*{Acknowledgements}
This work was supported by the Australian Government through the Australian Research Council Centre of Excellence grant (CE200100010). This research was supported by the Commonwealth through an Australian Government Research Training Program Scholarship [DOI: https://doi.org/10.82133/C42F-K220]

This work was performed in part at the Melbourne Centre for Nanofabrication (MCN) in the Victorian Node of the Australian National Fabrication Facility (ANFF), supported by an ANFF/MCN Technology Ambassador Fellowship, and in part at the Materials Characterisation and Fabrication Platform (MCFP) at the University of Melbourne and the Victorian Node of the ANFF.

The authors would like to thank Dr Nitu Syed for performing the ellipsometry of the resist AZ5214E and Haiwei Wang for their useful discussions.

\section*{Conflict of Interest}
The authors have no conflicts to disclose.

\section*{Author Contributions}

L.C, L.W and A.R conceptualised the project. L.C. designed, did simulations on and fabricated the device. A preliminary version of the device was fabricated with assistance from L.W. Y.X built the BFP spectroscopy setup and L.C and Y.X performed spectroscopy on the metasurface. L.C fabricated and characterised the greyscale phase objects. Imaging experiments on the greyscale objects and HeLa cells were performed by L.C. Imaging experiments on the tissue samples were performed by L.C with initial assistance from C.U.P. S.M prepared the HeLa cell samples. The tissue sample was prepared by D.J.B. An initial tissue sample was prepared by D.J.B and C.U.P. K.L.G provided analysis of the ovarian cancer tissue sample. L.C and A.R developed the manuscript. All authors provided input on the manuscript. E.H, K.L.G and A.R supervised the project.

\section*{Data Availability}
Data underlying the results presented in this paper are not publicly available at this time but may be obtained from the authors upon reasonable request.

\printbibliography

\end{refsection}

\begin{refsection}
    \pagebreak
\begin{center}
\textbf{\Large Supplemental Materials}
\end{center}

\setcounter{equation}{0}
\setcounter{figure}{0}
\setcounter{table}{0}
\setcounter{page}{1}
\setcounter{section}{0}

\makeatletter
\renewcommand{\theequation}{S\arabic{equation}}
\renewcommand{\thefigure}{S\arabic{figure}}
\renewcommand{\thesection}{S\arabic{section}}

\section{Topology Optimisation Algorithm}

\begin{figure}[ht]
    \centering
    \includegraphics[width=\linewidth]{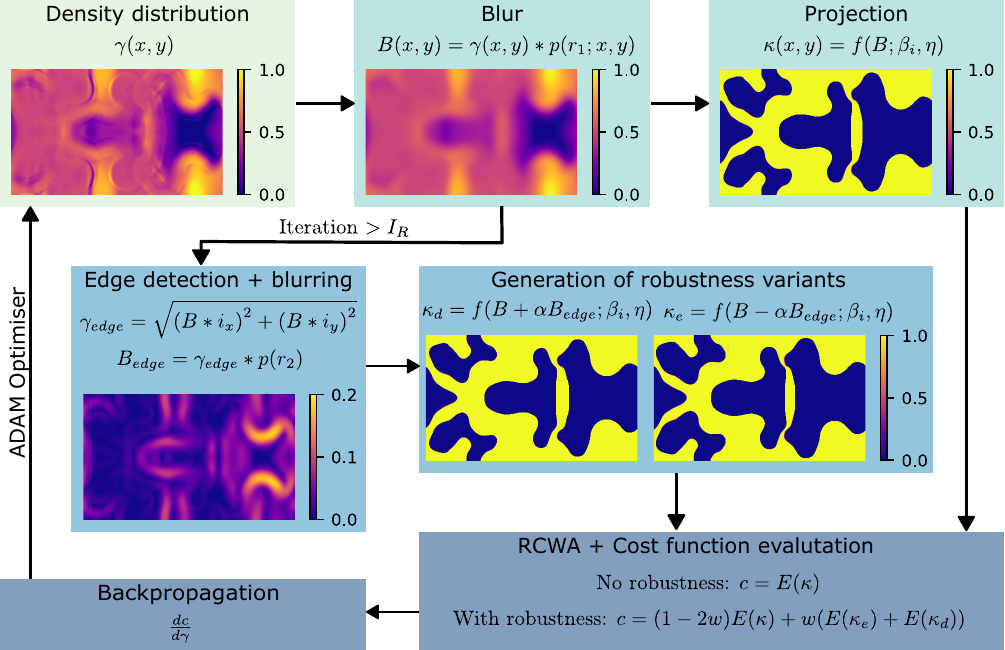}
    \caption{Flowchart depicting the optimisation loop. During each iteration various filtering and image processing operations are performed to generate robust designs.}
    \label{fig:optim_flowchart}
\end{figure}

A flowchart depicting the topology optimisation algorithm employed in this work is given in figure \ref{fig:optim_flowchart} and the full list of hyperparameters chosen for the optimisation are given in table \ref{tab:optimisation parameters}. There are a variety of different filtering steps required to produce designs that are both binary (composed of material or air), with suitable feature sizes and robust to fabrication defects.

The density distribution $\gamma (x, y)$ is the quantity that is optimised during the algorithm, representing the spatial distribution of material. We optimise on an $n_x$ $\times$ $n_y$ pixel grid and $\gamma$ is initialised with random values between 0 and 1. Before the optimisation loop commences, we enforce symmetry in $\gamma$ about the $x$-axis and blur it.

After each iteration we perform a blurring step to remove single pixel features and enforce a minimum feature size. This operation is a convolution, $B = \gamma * p(r_1)$, where $B$ is the blurred density and the blurring kernel, $p(r_1)$, is a cone with radius $r_1$ and $*$ denotes convolution.
\begin{equation}
    p(x, y; r) = 
    \begin{cases} 
        \frac{r - \sqrt{x^2 + y^2}}{1/3 \pi r^3} & \sqrt{x^2 + y^2} < r \\
        0 & otherwise 
    \end{cases}
    \label{eq:top_opt_blur}
\end{equation}

A projection function is used to enforce binarisation during the optimisation. As the number of iterations increases, the projection function approaches a step function which creates a density that is binary. The projection function, $f(A; \beta, \eta)$, and blurring kernel, $p(x, y; r)$ are taken from Wang et al. \cite{Wang2010topopt}. 
\begin{equation}
    f(A; \beta, \eta) = \frac{\tanh{(\beta \eta)} + \tanh{(\beta(A - \eta))}}{\tanh{(\beta \eta)} + \tanh{(\beta(1 - \eta))}}
    \label{eq:top_opt_proj}
\end{equation}
The parameter $\beta$ controls the strength of the binarisation in the projection function and $\eta$ controls the threshold value, we set $\eta = 0.5$. Every $N_\beta$ iterations we increase the value of $\beta$ by multiplying it by $F_\beta$. The projection step, $\kappa = f(B; \beta_i, \eta)$ produces the regular metasurface design, where the variable $\kappa$ directly corresponds to the permittivity distribution of the unit cell, via,
\begin{equation}
    \epsilon(x,y) = \epsilon_{\mathrm{si}} + (\epsilon_{\mathrm{air}} - \epsilon_{\mathrm{si}}) (1 - \kappa(x, y) )
\end{equation}
where $\epsilon (x,y)$ is the permittivity at a given point in the unit cell, $\epsilon_{\mathrm{air}}$ is the permittivity of air and $\epsilon_{\mathrm{si}}$ is the permittivity of amorphous silicon. 

During the first $I_R$ iterations, we consider only the regular metasurface design, not incorporating the effect of fabrication defects. Topology optimisation can produce designs that are very sensitive to small changes in the geometry. We wish to produce devices that are tolerant to fabrication defects, in particular robustness to under or over etched designs. While many algorithms exist for generating these robustness variants \cite{Wang2010topopt}, we use a method inspired by \cite{Kim2024SI}, where edge detection techniques are used to dilate or shrink the boundary of the structure.

After iteration $I_R$, during each optimisation iteration we generate, in addition to the regular design, the eroded and dilated variants $\kappa_e$ and $\kappa_d$ respectively. First we use the Prewitt operators to perform edge detection on the blurred density, $B$. The edge image, $\gamma_{\mathrm{edge}}$ is built from convolutions between the blurred density and the Prewitt operators, $i_x$ and $i_y$.
\begin{equation}
    \gamma_{\mathrm{edge}} = \sqrt{(B * i_x)^2 + (B*i_y)^2}
\end{equation}
\begin{align}
    i_x &= \begin{bmatrix}
        1 & 0 & -1 \\
        1 & 0 & -1 \\
        1 & 0 & -1 
    \end{bmatrix} \\
    i_y &= \begin{bmatrix}
        1 & 1 & 1 \\
        0 & 0 & 0 \\
        -1 & -1 & -1 
    \end{bmatrix}
\end{align}
Next, the edge image is blurred with the same blurring kernel, but with a radius of $r_2$. Changing the value of $r_2$ changes by how far the edges are eroded or dilated. From the blurred edges, $B_{\mathrm{edge}}$ and the blurred density, $B$, we generate and project the two robustness variants.
\begin{align}
    \kappa_e &= f(B - \alpha B_{\mathrm{edge}}; \beta_i, \eta) \\
    \kappa_e &= f(B - \alpha B_{\mathrm{edge}}; \beta_i, \eta,
\end{align}
where $\alpha$ is a weighting factor.

Once the designs are generated, we perform an RCWA simulation (TORCWA \cite{torcwaSI}) and evaluate the S-parameters as a function of angle of incidence. During the RCWA simulation we use $(2M+1) \times (2N+1)$ orders in the Fourier expansion. 

The cost function that we wish to minimise, $E(A)$, is the root mean square error between a target optical transfer function and the OTF of the current design, $A$.
\begin{equation}
    E(A) = \sqrt{\frac{1}{N_\mathrm{angle} N_\mathrm{pol}} \sum_j\sum_k \left( t_{j, k} - T_{j, k} \right)^2}
\end{equation}
where $N_\mathrm{angle}$ is the number of angles at which we evaluate the OTF, $N_\mathrm{pol}$ is the number of polarisations we consider, the sum over $j$ is the sum over the polarisation, the sum over $k$ is the sum over angle, $t_{j, k}$ is the metasurface 0th order amplitude transmission and $T_{j, k}$ is the target amplitude transmission for polarisation $j$ and angle $k$. Prior to iteration $I_R$, the cost function is simply $E(\kappa)$. However, when considering robustness, the cost function becomes a weighted average of the two robustness variants with the regular variant, given by,
\begin{equation}
    c = (1-2w) E(\kappa) + wE(\kappa_e) + wE(\kappa_d),
\end{equation}
where $w$ is the robustness weighting.

After the cost function is evaluated we calculate its gradient with respect to $\gamma$ using automatic differentiation. Then pass this gradient into the gradient based optimisation algorithm ADAM \cite{ADAM} to update $\gamma$. ADAM is a gradient based algorithm that uses estimates of the lower-order moments to update the parameters each iteration. There are 4 hyperparameters, $\alpha_{\mathrm{step}}$, which represents the maximum change to the density each iteration, $\beta_1$, the decay rate of the first moment, $\beta_2$, the decay rate of the second moment, and $\epsilon_z$ which prevents division by 0 errors. 

We then impose the symmetry about the $x$-axis constraint by applying the operation $\gamma_{sym} (x, y) = \frac{\gamma(x, y) + \gamma(x, -y)}{2}$. The optimisation is run for $N_{\mathrm{iter}}$ optimisations and the final design is given by $\kappa(x, y)$.

The full list of hyperparameters used during the optimisation are given below in table \ref{tab:optimisation parameters}.

\begin{table}[H]
\begin{center}
    \begin{tabular}{cc|p{12cm}}
         Parameter & Value & Description\\ \hline
         $n_x$ & 425 & Number of pixels in $x$-direction in final design \\
         $n_y$ & 250 & Number of pixels in $y$-direction in final design\\
         $\eta$ & 0.5 & Threshold value\\
         $N_\beta$ & 30 & How many iterations before $\beta$ is increased\\
         $ F_\beta$ & 1.2 & $\beta$ increase factor\\
         $r_1$ & 40 nm & Blurring kernel radius\\
         $I_R$ & 350 & Number of iterations without robustness\\
         $r_2$ & 8 nm & Edge blurring kernel radius\\
         $\alpha$ & 0.25 & Edge weighting in robustness variant generation\\
         $w$ & 0.25 & Weighting of robustness in cost function\\
         $M$ & 16 & $2M+1$ Fourier coefficients in RCWA expansion in the $x$-direction\\
         $N$ & 10 & $2N+1$ Fourier coefficients in RCWA expansion in the $y$-direction\\
         $N_{\mathrm{iter}}$ & 1500 & Total number of iterations\\
         $\alpha_{\mathrm{step}}$ & 0.01 & Maximum step size in ADAM optimisation\\
         $\beta_1$ & 0.9 & ADAM decay of 1st moment \\
         $\beta_2$ & 0.999 & ADAM decay of 2nd moment\\
         $\epsilon_z$ & $10^{-6}$ & ADAM prevent division by 0\\
    \end{tabular}
    \caption{Optimisation hyperparameters}
    \label{tab:optimisation parameters}
\end{center}
\end{table}

\section{Robustness}

\begin{figure}[H]
    \centering
    \includegraphics[width=\linewidth]{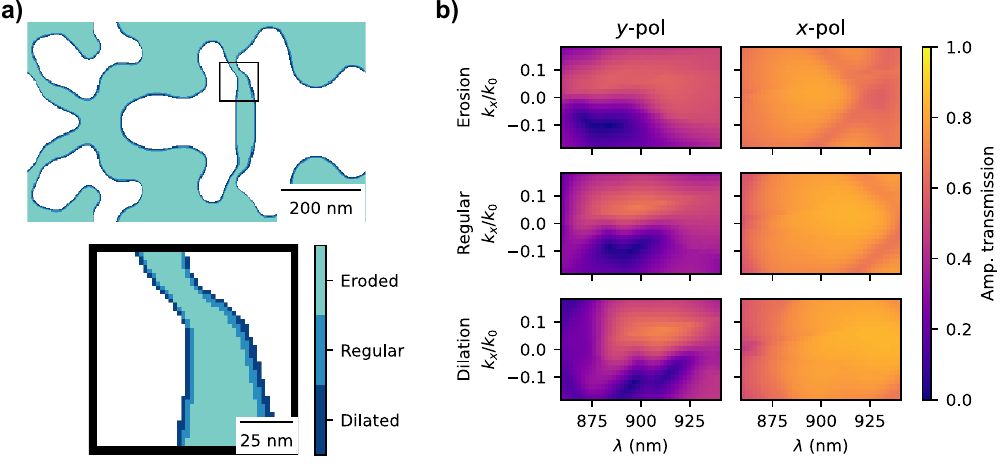}
    \caption{Effect of fabrication defects on the OTF of the metasurface. \textbf{a)} Robustness variants generated throughout the optimisation algorithm overlayed on top of each other. Coloured region represented the silicon region. \textbf{b)} OTF for each metasurface variant as a function of wavelength computed using RCWA.}
    \label{fig:robustness}
\end{figure}

In order to analyse the robustness of the structure to the fabrication defects considered in our optimisation algorithm, we compute the OTF as a function of wavelength for the eroded, dilated and regular variants of the metasurface, the result of which is shown in figure \ref{fig:robustness}.

Figure \ref{fig:robustness}a shows the 3 variants which are generated during the last iteration of the optimisation algorithm overlaid on top of each other. The eroded and dilated versions of the structure represent an erosion or dilation of the structure by about 4 nm. The extent of this variation is of the same order of magnitude, but larger than, the smallest beam sizes of the electron beam lithography process, so is suitable to model realistic changes in the geometry that may occur. When analysing the OTF of these variants (figure \ref{fig:robustness}b), there is some blue and red-shifting of the linear region of the OTF. The eroded variant demonstrates blue-shifting, the optimal wavelength shifts to 880 nm, whereas the dilated structure exhibits redshifting moving to 910 nm. The operational wavelength range of this metasurface is relatively broad, with a monotonically increasing $y$-polarised OTF existing for all variants over a 30 nm range. The $x$-polarised OTF has very minimal features. Eroding and dilating the structure has a minimal effect on the brightfield imaging mode of this metasurface.

\begin{figure}
    \centering
    \includegraphics[width=0.5\linewidth]{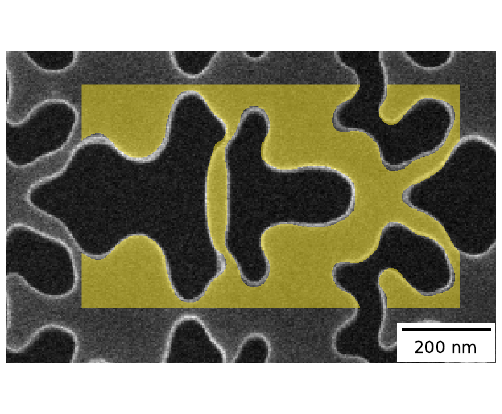}
    \caption{Comparison between the fabricated metasurface geometry and the designed geometry. Yellow region is the unit cell of the designed metasurface.}
    \label{fig:rob_comp}
\end{figure}

Figure \ref{fig:rob_comp} compares the fabricated metasurface geometry with the designed geometry. Generally, the silicon region of the fabricated metasurface is slightly dilated compared to the ideal design, but this dilation is not-uniform in contrast to our simulated robustness analysis. In some regions, the design and fabricated metasurface match quite well and in other areas there is a small offset. The combination of non-uniform dilation and potentially small-mismatches in refractive index data of the deposited aSi films and the data used in simulations could result in the small blue-shift we see in experiments.

\section{Full Wave FEM Modelling}

We model the optical properties of the metasurface using the finite element method (FEM) as implemented in COMSOL Multiphysics v6.3. To import the metasurface geometry into COMSOL, we convert the pixel design produced by the topology optimisation algorithm into a GDS file where each pixel is translated into a square. We convert the GDS to a DXF which can be imported into COMSOL. We use periodic boundary conditions with Floquet periodicity to model an infinite array of the singular unit cell. The sub- and superstrate are modelled as semi-infinite. The height of the modelled region is 4.3 \um. Periodic port boundary conditions are used at the top of the superstrate and bottom of the substrate, with wave excitation at the superstrate port. In the superstate, we consider 4 diffracted orders and in the substrate we consider 10 diffracted orders. The diffraction efficiency is measured in the substrate.

\section{Additional Simulations and Characterisation}

\subsection{Phase Response of the Metasurface}

The phase response of the metasurface is calculated using the FEM. We only analyse the phase of the 0th order which carries the processed image.

\begin{figure}[H]
    \centering
    \includegraphics[width=0.5\linewidth]{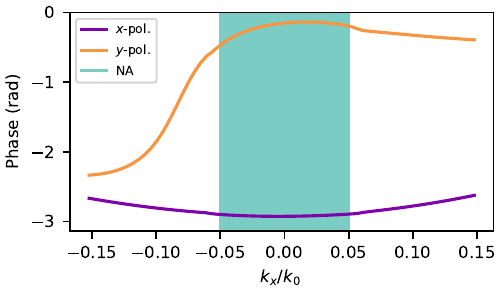}
    \caption{Phase transfer function of the metasurface}
    \label{fig:ms_phase}
\end{figure}

Within the processing NA of the metasurface, the phase transfer function is flat for both $x$- and $y$-polarisations. Outside the NA of 0.05, the $x$-polarised phase response remains reasonably flat but with a slight curve. The  $y$-polarised phase response, on the other hand, has a sharp dip for negative spatial frequencies. This rapid change to the phase is outside the processing NA of the metasurface so would have minimal effect on the imaging ability of our device. 

\subsection{Effect of Alumina Layer on Optical Performance}

\begin{figure}[H]
    \centering
    \includegraphics[width = \linewidth]{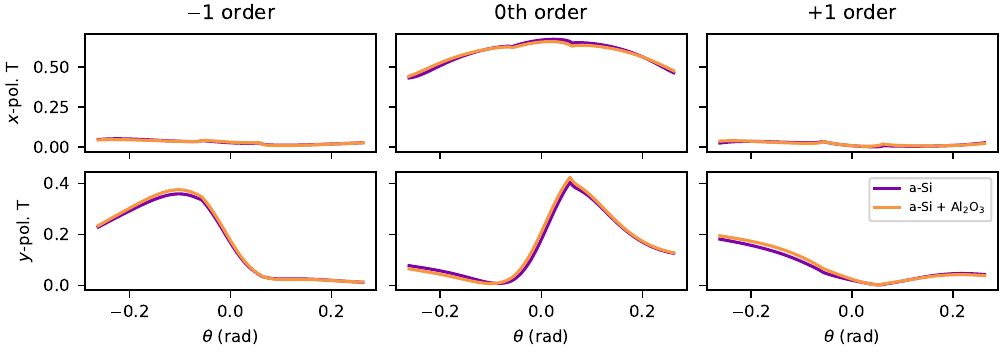}
    \caption{Diffraction efficiency as a function of angle of incidence calculated with the FEM when the alumina etch layer is included and not included. Diffraction efficiency for each diffracted order measured in the substrate.}
    \label{fig:etch_mask_included}
\end{figure}

We perform an additional FEM simulation to calculate the optical properties of the metasurface with the Al$_2$O$_3$ etch mask in place. The alumina layer has the same pattern as the underlying silicon, is 20 nm thick and we model it using optical constants from \cite{Boidin2016}. The diffraction efficiency as a function of spatial frequency at 900 nm is given in figure \ref{fig:etch_mask_included}.

As shown in figure \ref{fig:etch_mask_included}, the effect on the optical properties is very minor. The inclusion of the alumina increases the diffraction efficiency of the metasurface and creates greater contrast in the $y$-polarised OTF.

\subsection{2D Ray Tracing Simulations}

\begin{figure}[H]
    \centering
    \includegraphics[width=\linewidth]{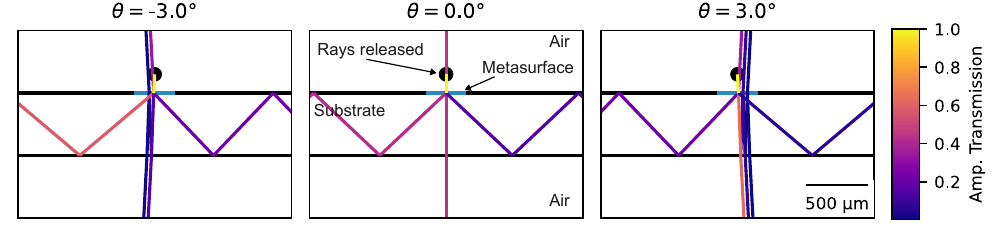}
    \caption{Ray tracing simulations}
    \label{fig:ray_tracing}
\end{figure}

To model the effect of the substrate on the diffracted orders, we perform 2D ray tracing simulations in COMSOL Multiplysics v6.3. We model the metasurface as a 300 \um~long grating on a 500 \um~tall silica substrate (optical constants \cite{Malitson1965sio2}. We model a 4 mm $\times$ 1.5 mm area. At the boundary of our model we use a wall boundary condition to freeze rays leaving the model. We release the rays 150 \um directly above the metasurface with a wavelength of 900 nm. The metasurface is modelled as a grating with diffraction efficiencies as a function of angle of incidence calculated from the full wave electromagnetic simulations. One thing to note is that the ray tracing and full wave electromagnetic solvers use different conventions in the labelling of the diffracted orders. We propagate the rays through the geometry for 10 ns. The ray tracing solution for different angles of incidence is given in figure \ref{fig:ray_tracing}.
We can see that only the 0th order ray emerges from the substrate and the $+1$ and $-1$ diffracted orders totally internally reflect inside the substrate. At $-3\degree$ the energy is concentrated in the $-1$ order. As the angle increases, the diffraction efficiency into this order decreases and the intensity of the transmitted 0th order increases. There are additional rays leaving the substrate with very low intensity that results from reflections inside the substrate. These have very low intensity and should have only a marginal effect on imaging performance.

\subsection{2D Optical Transfer Function}

\begin{figure}[H]
    \centering
    \includegraphics[width=\linewidth]{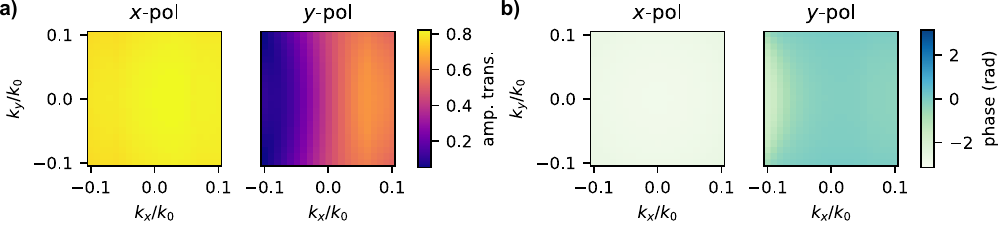}
    \caption{Simulated 2D OTF at 900 nm. \textbf{a)} amplitude transfer function, \textbf{b)} phase transfer function.}
    \label{fig:sim_2d_otf}
\end{figure}

We calculate the full 2D OTF for $x$- and $y$-polarised light with the FEM. The amplitude and phase of the OTF are presented in figure \ref{fig:sim_2d_otf}.
As desired, there is only a small variation in the OTF in both the amplitude and phase along the $k_y$ direction. Variation in the $k_y$ direction is symmetric, due to enforcing the symmetry requirement about the $x$-axis in the metasurface design. Furthermore, the small amount of variation in the OTF occurs outside the design NA of 0.05.

\begin{figure}[H]
    \centering
    \includegraphics[width=100mm]{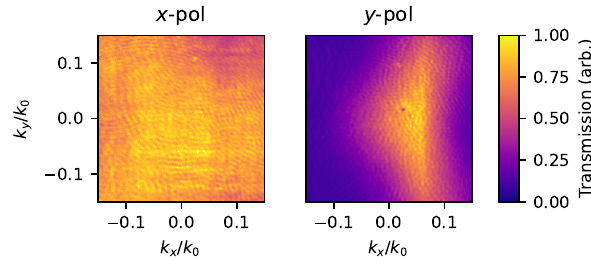}
    \caption{Measured 2D back-focal plane of the metasurface at 886 nm}
    \label{fig:meas_bfp}
\end{figure}

We validate the 2D OTF by imaging the back-focal plane of the fabricated metasurface at 886 nm. The transmission as a function of spatial frequency is given in figure \ref{fig:meas_bfp}.
The transmission in the $k_y$ direction is symmetric as designed. Compared to simulations, there is more variation along the $k_y$ direction, but within the 0.05 NA, this variation is small and the impact on the computational ability of the metasurface is small. At $k_x/k_0 = 0.05$, there is a sharp dip in the transmission, which is reflected in simulations and in the spectrally resolved spatial frequency response measurement presented in the main text.

\section{Phase Object Fabrication}

\begin{figure}[ht]
    \centering
    \includegraphics[width=0.5\linewidth]{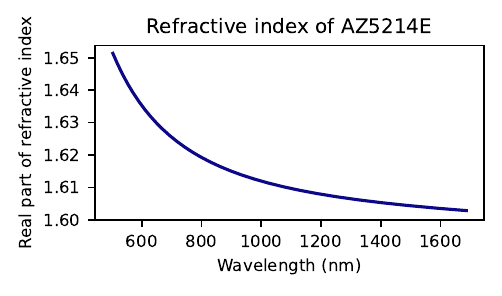}
    \caption{Refractive index of 1.4 \um of photoresist AZ5214E on silicon substrate. Imaginary part of refractive index is below the minimum measurable value for this wavelength range. Measured using ellipsometry by Dr. Nitu Syed.}
    \label{fig:nk}
\end{figure}

Phase objects are fabricated with greyscale photolithography out of resist AZ5214E. Greyscale lithography permits us to fabricate 2.5D topographies out of a positive photoresist. If the photoresist is transparent, then by varying the thickness of the resist at each point, we can control the phase. The phase imparted to the light, $\phi (x, y)$, passing through the transparent object is given by,
\begin{equation}
    \phi (x, y) = \frac{2\pi \Delta n}{\lambda} T(x, y),
\end{equation}
where $\Delta n$ is the difference in refractive index between the resist and air, $\lambda$ is the free space wavelength and $T(x, y)$ is the thickness profile of the resist. The refractive index of the resist was measured using ellipsometry on a silicon wafer substrate and is given in figure \ref{fig:nk}.

The resist, AZ5214E, is highly transparent in the near infrared with a refractive index of approximately 1.6. Using our greyscale lithography recipe, we can create topographies with a maximum height variation of 857 nm, in order to extend this range, a thicker layer of resist can be spin-coated and a larger dose can be used. This 857 nm depth variation corresponds to a maximum phase excursion of approximately 3.6 rad with 900 nm illumination.

\begin{figure}[H]
    \centering
    \includegraphics[width = \linewidth]{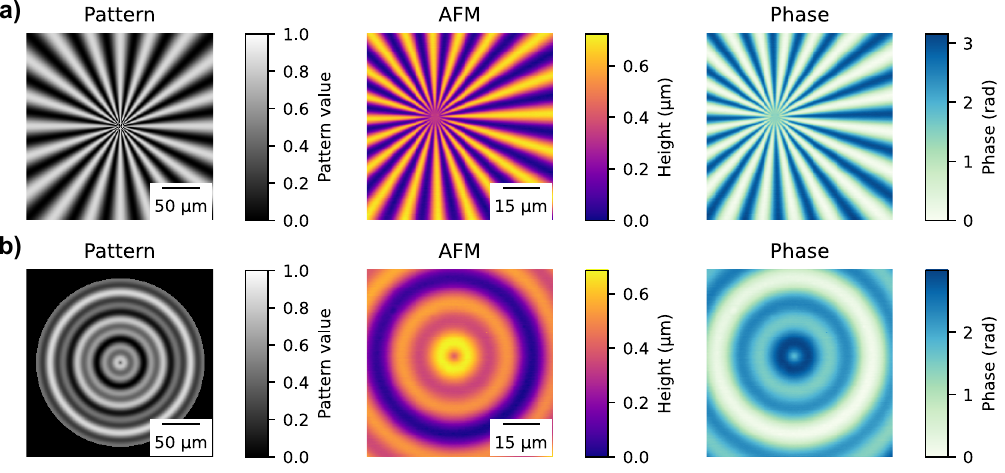}
    \caption{Phase objects used for imaging experiments. In the pattern, a value of 0 corresponds to the bottom of the photoresist and a value of 1 corresponds to the top of the photoresist. Phase of the object is calculated at 886 nm. \textbf{a)} Phase spoke target with a $\pi$ phase excursion, \textbf{b)}, phase ring structure.}
    \label{fig:phase_objects}
\end{figure}

The greyscale lithography pattern is a greyscale image, where 0 (black), corresponds to the lowest height, and 1 (white) corresponds to the maximum height of 857 nm. In the main text we demonstrate the phase imaging using two fabricated phase objects. The phase of these patterns, $\phi(x,y)$, is defined by the equations below, with $x$ and $y$ in \um.
\begin{equation}
    \phi_{\mathrm{spoke}}(x, y) = \frac{\pi}{2} \left[ \cos{\left( n \tan^{-1}{\left( \frac{y}{x} \right)}\right)} + 1\right]
    \label{eq:phase_spoke}
\end{equation}
\begin{equation}
    \phi_{\mathrm{bullseye}}(x,y) = 
    \begin{cases}
        \frac{\pi}{2} \left[ \cos{\left( \frac{4\pi}{125} \sqrt{x^2 + y^2} \right) \sin{\left(\frac{4\pi}{50} \sqrt{x^2 + y^2} \right)}} + 1 \right] & \sqrt{x^2 + y^2} \leq 125 \text{ {\textmu}m}\\
        0 & \sqrt{x^2 + y^2} > 125 \text{ {\textmu}m}\\
    \end{cases} 
    \label{eq:bullseye}
\end{equation}
The spoke target is generated with equation \ref{eq:phase_spoke} using $n=20$ and the bullseye target is generated with equation \ref{eq:bullseye}. The pattern defining these objects, the AFM of the resulting fabricated structure and the phase derived from the AFM are given in figure \ref{fig:phase_objects}.

\begin{figure}[H]
    \centering
    \includegraphics[width=\linewidth]{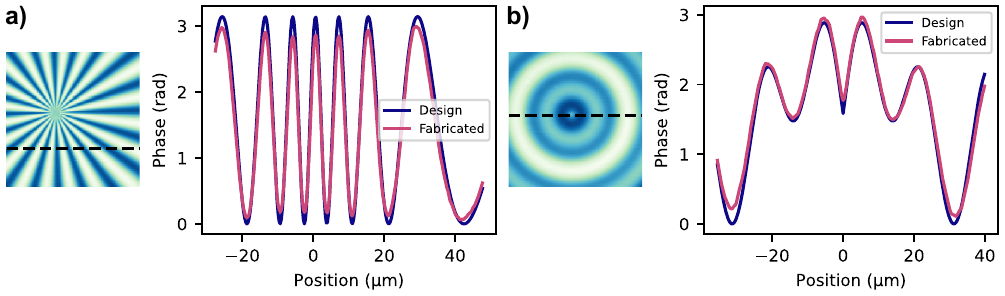}
    \caption{Line profile through the fabricated phase objects comparing the fabricated phase structure to the design phase profile. \textbf{a)} Spoke target, \textbf{b)} "bullseye`` phase target.}
    \label{fig:phase_profile}
\end{figure}

In figure \ref{fig:phase_profile} we show the line profiles of the fabricated phase test targets compared to the design phase profile. The shape of the phase profiles agree quite well, however the fabricated targets have a slightly lower maximum phase excursion than desired. In particular, the spoke target has a maximum phase excursion of $0.934 \pi$ radians and the "bullseye`` target has a maximum phase excursion of $0.993 \pi$. In our comparison between the recovered and ground truth phase gradients of these phase objects, we multiply the ground truth phase so that the phase excursion matches the fabricated samples. 

\section{Phase Gradient Recovery Fitting Parameters}

To remove the tilt during the phase gradient recovery, we use a very simple optimisation procedure to calculate a value for $\beta$, representing the transmission of the metasurface at normal incidence. To begin, we set $\beta=0.706$, the value obtained from spectroscopy. We then calculate the phase gradients, $d\phi/dx$ from the $x$ and $y$ polarised images. If the average gradient is larger than 0.01 rad/\um, then we set $\beta_{new} = \beta_{old} + \frac{1}{2}\mathrm{average}\left( d\phi/dx \right)$. We then repeat until until the average phase gradient is less than 0.01 rad/\um. Since we align the metasurface to be perpendicular to the beam, the amount of tilt is small and the fitting converges quickly. 

In table \ref{tab:fit_params}, we include the fitting parameters used in the phase gradient recovery in the main text for the tilt-removal process and the tilt that is implied by this fitting parameter.

\begin{table}[H]
\begin{center}
    \begin{tabular}{c|c|c}
        Image & $\beta$ & Implied tilt\\
        \hline
        Spoke & 0.700 & $-0.12\degree$\\
        Bullseye & 0.719 & $+0.22 \degree$\\
        HeLa & 0.625 &  $-1.3 \degree$\\
        Ovarian cancer tissue & 0.789 & $+1.3 \degree$\\
    \end{tabular}
    \caption{Fitting parameters used for phase gradient recovery}
    \label{tab:fit_params}
\end{center}
\end{table}

Not all the tilt can be attributed to the metasurface. The biological imaging requires a coverslip with water to be placed on top of the sample. The coverslip itself could be at a slight angle, or the sample could be tilted with respect to the optical axis.

\section{Optical Characterisation}

\begin{figure}[H]
    \centering
    \includegraphics[width=0.5\linewidth]{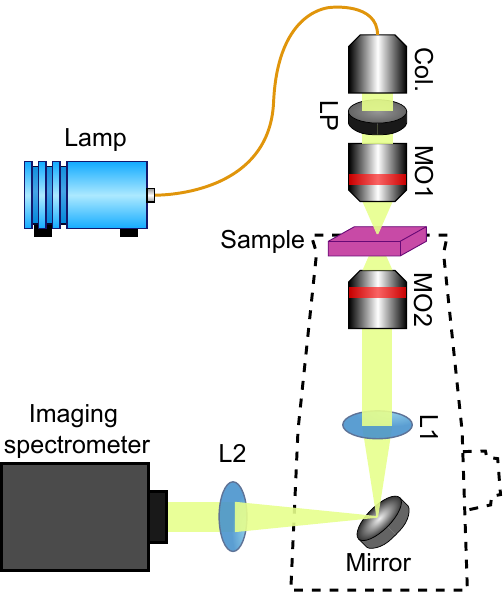}
    \caption{Back focal plane spectrometry setup}
    \label{fig:BFP_spec}
\end{figure}

To measure the 1D OTF and spectrum of the metasurface simultaneously, we use the back focal plane (BFP) spectrometry setup shown in figure \ref{fig:BFP_spec}. This setup is built into an inverted microscope (Nikon Eclipse Ti-U). Light from a halogen lamp (Mikropack HL-2000 FHSA) is collimated (ThorLabs TC25FFFC-633) and then polarised (ThorLabs LPVIS100-MP2). The light travels through MO1 (Leica N Plan 5$\times$) which creates a cone of light. The sample is placed at the focal plane of the objective and is imaged using another objective MO2 (Nikon Plan Fluor $10\times$). The numerical aperture of the back focal plane is defined by MO1, which has an NA of 0.12.

The light travels through the microscope towards L1, the internal microscope tube lens, before bouncing off a mirror and heading towards the imaging spectrometer. Lens L2 (200 mm ThorLabs LA1708-A) is used to image the back focal plane of the objective MO2 onto the sensor of the spectrometer (Shamrock 303i spectrograph equipped with an Andor iDus 420A CCD). The spectrometer features an adjustable slit, which is used to take a 1D slice of the full 2D BFP and a blazed grating is used to spectrally resolve the OTF. We set the slit width to 260 \um.

\begin{figure}[H]
    \centering
    \includegraphics[width=\linewidth]{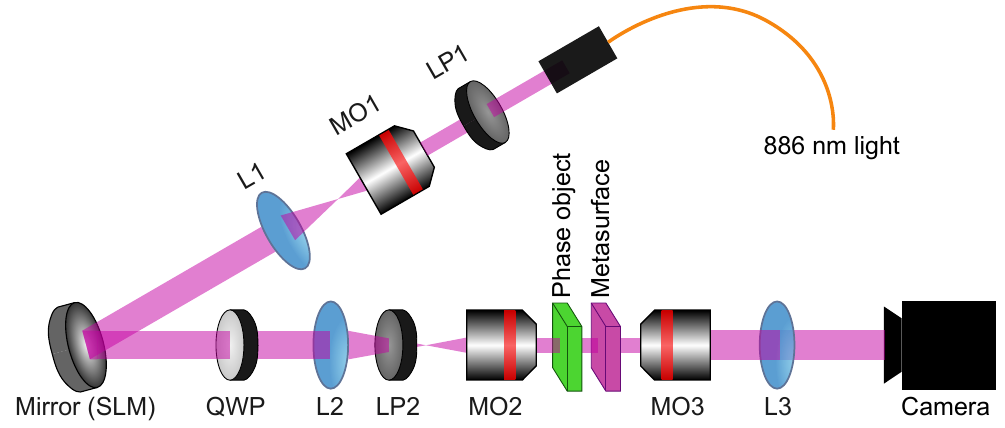}
    \caption{Setup used for imaging of the phase targets.}
    \label{fig:phase_imaging_setup}
\end{figure}

We perform the image processing experiments of various samples in a custom built microscope setup, given in figure \ref{fig:phase_imaging_setup}. The section of the setup before the SLM collimates and resizes the beam. Light is injected into the optical system from an NKT Photonics SuperK Compact with a SuperK SELECT acousto-optic filter, LP1 is a linear polariser (ThorLabs LPNIR050-MP2). The beam is then resized with MO1 (Olympus Plan N 20 $\times$) and L1 (125 mm ThorLabs LA1986B). The light bounces off an SLM, which is turned off and acts as a mirror. 

We adjust the polarisation of the beam and resize it so that only a small section of the sample is illuminated. A quarter wave plate (ThorLabs AQWP10M-580) is used to create circularly polarised light. L2 (150 mm ThorLabs LA1433B) and MO2 (Olympus Plan N 20 $\times$) are used to shrink the beam, ensuring that the beam is collimated after leaving the objective. LP2 (ThorLabs LPVIS050-MP) is used to select the linear polarisation state that we want.

The phase object is imaged onto the camera (Andor Zyla) with objective MO3 (Nikon LWD 20 $\times$) and tube lens L3 (200 mm ThorLabs AC254-200-A-ML). 

\begin{figure}[H]
    \centering
    \includegraphics[width=\linewidth]{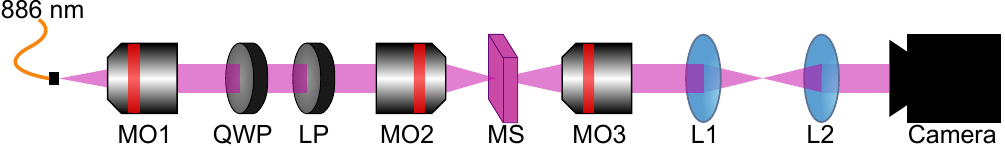}
    \caption{2D back-focal plane imaging setup.}
    \label{fig:2D_BFP_setup}
\end{figure}

To measure the 2D spatial frequency response of the metasurface we image the back-focal plane using the setup presented in figure \ref{fig:2D_BFP_setup}. Light is coupled into the system with an NKT Photonics SuperK Compact with a SuperK SELECT acousto-optic filter. The light is collimated wth MO1 (Olympus PlanN 4$\times$), we polarise the light with a quarter wave plate, QWP (Thorlabs AQWP05M-600) and linear polariser, LP (Thorlabs LPNIR 100). We illuminate the metasurface (MS) with a cone of light using MO2 (Olympus UPlanFl 20$\times$) and collect the light with MO3 (Olympus UPlanFl 10$\times$). Lenses L1 (300 mm Thorlabs LA1484-B) and L2 (200 mm Thorlabs AC254-200-A-ML) are used to image the back focal plane of MO3 onto the camera (Thorlabs CS505MUP1).

\section{Biological Sample Images}

Here we present the biological images presented in the main text but without the averaging over the metasurface position. The same details as the images in the main text are present but permeated by diffraction from the finite size of the metasurface and from defects in the metasurface. The pseudo-3D effect that we expect from these samples is still visible and we can still infer information about the cells and textures of the samples. Compared to the greyscale phase objects, the effect of diffraction is greater since there is an additional coverslip present and the metasurface cannot be positioned as close to the sample.

\begin{figure}[H]
    \centering
    \includegraphics[width=\linewidth]{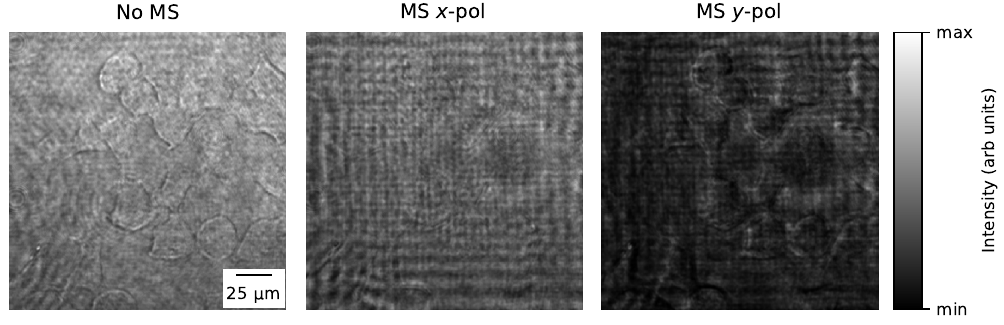}
    \caption{Image of the HeLa cells without averaging over the metasurface position}
    \label{fig:hela_no_ave}
\end{figure}

\begin{figure}[H]
    \centering
    \includegraphics[width=\linewidth]{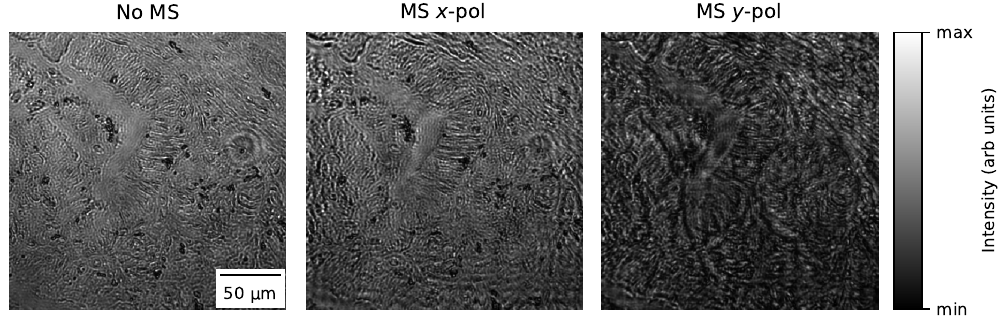}
    \caption{Image of the tissue sample without averaging the metasurface position}
    \label{fig:tissue_no_ave}
\end{figure}

Subtracting off these artefacts by either keeping the metasurface stationary and taking an image without the sample or by averaging over the metasurface position leads to a much clearer image and better phase gradient recovery. 

\section{Finite Size of the Metasurface}

We use the angular spectrum propagation method to analyse the effect of diffraction from the metasurface edges resulting from defocus and the finite size of the metasurface. The metasurface is modelled as a square aperture with side-length $l$. The background is 100\% transmitting and the metasurface is 50\% transmitting. The metasurface lies in the $x$, $y$ plane. The field from the metasurface plane is given by $U_0 (x, y)$, where,
\begin{equation}
    U_0(x, y) =
    \begin{cases}
    0.5, & |x|\leq l/2 ~\mathrm{and}~ |y| \leq l/2 \\
    1, & \mathrm{otherwise}.
\end{cases}
\end{equation}
The field along the $z$-axis measured at a distance $\Delta z$ away from the metasurface (representing the distance between metasurface and phase object) is given by $U(x,y)$.
\begin{equation}
    U(x, y) =\mathcal{F}^{-1}\left[ \mathcal{F}\left[U_0 (x, y) \right] e^{ik_z \Delta z}\right],
\end{equation}
where $\mathcal{F}$ and $\mathcal{F}^{-1}$ represent the Fourier and inverse Fourier transform respectively and $k_z$ is the wavevector in the $z$-direction, where $k_z = \sqrt{k_0^2 - k_x^2-ky^2}$, $k_0 = 2\pi/\lambda$, $\lambda$ is the free-space wavelength and $k_x$ and $k_y$ are the spatial frequencies. The intensity is given by $I(x, y) = |U(x, y)|^2$ and we define the variation in intensity as $\Delta I = \max(I(x, y)) - \min(I(x, y))$. We wish to find an area over which the variation in intensity is smaller than some threshold. This area represents the field of view of the metasurface for a given acceptable amount of diffraction.

\begin{table}[H]
    \centering
    \begin{tabular}{c|c|c|c}
        Metasurface-object distance & $\Delta I = 0.1$ & $\Delta I =0.05$ & $\Delta I = 0.02$ \\
        \hline
        50 \um & $257 \times 257$ & $255 \times 255$ & $236 \times 236$ \\
        100 \um & $226 \times 226$ & $198 \times 198$ & $191 \times 191$ \\
        200 \um & $222 \times 222$ & $105 \times 105$ & $87 \times 87$ \\
        300 \um & $195 \times 195$ & $22 \times 22$ & $17 \times 17$ \\
    \end{tabular}
    \caption{Diffraction minimised field of view (in \um) for defocused 300 \um~$\times$ 300 \um~metasurface.}
    \label{tab:300um_finite_size}
\end{table}

As expected, with the metasurface positioned as close as possible to the sample, the effects of diffraction are minimal and the field of view over which diffraction is minimised is large. For the imaging experiments on biological samples, the metasurface and sample are separated by a coverslip which is approximately 200 \um~thick, which results in the strong appearance of the fringes in figures \ref{fig:hela_no_ave} and \ref{fig:tissue_no_ave}. With 300 \um~ of defocus, there is only a $22 \times 22$ \um$^2$ region where diffraction is minimised such that $\Delta I \leq 0.05$. As demonstrated in the main text, one method of combatting the finite size of the metasurface is to capture several images with the metasurface slightly offset and average the image. For practical applications, the metasurface can be made larger

\begin{table}[H]
    \centering
    \begin{tabular}{c|c|c|c}
        Metasurface-object distance & $\Delta I = 0.1$ & $\Delta I =0.05$ & $\Delta I = 0.02$ \\
        \hline
        50 \um & $958 \times 958$ & $957 \times 957$ & $937 \times 937$ \\
        100 \um & $927 \times 927$ & $900 \times 900$ & $892 \times 892$ \\
        200 \um & $923 \times 923$ & $804 \times 804$ & $790 \times 790$ \\
        300 \um & $896 \times 896$ & $728 \times 728$ & $689 \times 689$ \\
    \end{tabular}
    \caption{Diffraction minimised field of view (in \um) for defocused 1 mm $\times$ 1 mm metasurface.}
    \label{tab:1mm_finite_size}
\end{table}

For example, a $1\times 1$ mm$^2$ metasurface can achieve a $689 \times 689$ \um$^2$ field of view with only $\Delta I = 0.02$.
    \printbibliography
\end{refsection}

\end{document}